%% file: journal_v4.tex
\documentclass[journal, onecolumn]{IEEEtran}

\usepackage{cite}
\usepackage{graphicx,amsmath,epsfig,times}
\usepackage{caption,subcaption}
\usepackage{amstext,amssymb,latexsym}
\usepackage{color}
\usepackage{enumitem}
\usepackage{bbm}
\usepackage{mathrsfs}
\input{defns}

\newtheorem{theorem}{Theorem}
\newtheorem{lemma}{Lemma}
\newtheorem{proposition}{Proposition}
\newtheorem{corollary}{Corollary}
\newtheorem{example}{Example}
\newtheorem{claim}{Claim}
\newtheorem{remark}{Remark}

\newcommand{\Ob}{\mathbf{0}}
\newcommand{\Fq}{\mathbb{F}_q}
\newcommand{\Ffi}{\mathbb{F}}
\newcommand{\GF}{\mathrm{GF}}
\newcommand{\Mc}{\mathcal{M}}
\newcommand{\Rrl}{\Rr_\mathrm{L}}
\newcommand{\tRr}{\tilde{\Rr}}
\newcommand{\tRrl}{\tilde{\Rr}_\mathrm{L}}
\newcommand{\Rrmac}{\Rr_\mathrm{MAC}}
\newcommand{\vf}{\varphi}
\newcommand{\sbq}{\subseteq} 

\newcommand{\Dr}{\mathscr{D}}

\allowdisplaybreaks

\IEEEoverridecommandlockouts
\begin{document}

\title{Homologous Codes for\\ Multiple Access Channels}

\author{\IEEEauthorblockN{Pinar Sen~\IEEEmembership{Student Member,~IEEE,}
and Young-Han Kim~\IEEEmembership{Fellow,~IEEE,}}
\thanks{This work was supported in part by the Electronics and Telecommunications Research Institute through Grant 17ZF1100 from the Korean Ministry of Science, ICT, and Future Planning.}
\thanks{This work was presented in part in the 2017 IEEE International Symposium of Information Theory.}
\thanks{Pinar Sen and Young-Han Kim are with the Department of Electrical and Computer Engineering, University of California, San Diego, La Jolla, CA 92093 USA (email: psen@ucsd.edu, yhk@ucsd.edu)}
}


\maketitle

\begin{abstract} 
Building on recent development by
Padakandla and Pradhan, and by Lim, Feng, Pastore, Nazer, and Gastpar,
this paper studies the potential of structured nested coset coding as a complete replacement for
random coding in network information theory. The roles of two techniques used
in nested coset coding to generate nonuniform codewords, namely, shaping and channel
transformation, are clarified and illustrated via the simple example of
the two-sender multiple access channel. While individually deficient, the optimal combination
of shaping and channel transformation is shown to achieve the same performance as
traditional random codes for the general two-sender multiple access channel. The achievability proof of the capacity region is extended to the multiple access channels with more than two senders, and with one or more receivers. A quantization argument consistent with the construction of nested coset codes is presented to prove achievability for their Gaussian counterparts. These results open up new possibilities of utilizing nested coset codes with the same generator matrix
for a broader class of applications.
\end{abstract}


%
\IEEEpeerreviewmaketitle

\section{Introduction}
Random independently and identically distributed (i.i.d.\@) code ensembles play
a fundamental role in network information theory,
with most existing coding schemes built on them; see, for example, \cite{El-Gamal--Kim2011,Cover--Thomas2006,Kramer2007}. As shown by the classical example by K\"orner and Marton \cite{Korner--Marton1979}, however, using the same code at multiple users can achieve strictly better performance for some communication problems. Recent studies illustrate the benefit of such \emph{structured coding} for computing linear combinations in \cite{Padakandla--Pradhan2013c,Nazer--Gastpar2011,Song--Devroye2013,Lim--Gastpar2016,Sen--Lim--Kim2018c,Sen--Lim--Kim2018s}, for the interference channels in~\cite{Padakandla--Pradhan2012,Ntranos2013,Ordentlich--Erez--Nazer2014,Padakandla--Pradhan2016}, and for the multiple access channels with state information in~\cite{Padakandla--Pradhan2013}. Consequently, there has been a flurry of research activities on structured coding in network information theory, facilitated in part by several standalone workshops and tutorials at major conferences by leading researchers.

Most of the existing results are based on lattice codes or linear codes on finite alphabets. Recently, Padakandla and Pradhan~\cite{Padakandla--Pradhan2013}  brought a new dimension to the arsenal of structured coding by developing nested coset codes for network information theory;
see also Miyake \cite{Miyake2010} for nested coset codes for point-to-point communication.
In these nested coset coding schemes, a coset code of a rate higher than the target is first generated randomly. A codeword of a desired property (such as type or joint type) is then selected from a subset (a coset of a subcode). This construction is reminiscent of the
multicoding scheme in Gelfand--Pinsker coding for channels with state and Marton coding for broadcast channels. But in a sense, nested coset coding is more fundamental in that the scheme at its core is relevant even for single-user communication. By a careful combination of individual and common parts of coset codes, the proposed coding scheme in \cite{Padakandla--Pradhan2013} achieves rates for multiple access channels (MACs) with state
beyond what can be achieved by existing random or structured coding schemes. The analysis of the scheme is performed by packing and covering lemmas developed again in \cite{Padakandla--Pradhan2013} that parallel such lemmas for random coding in \cite{El-Gamal--Kim2011}.

Recently, structured coding based on random nested coset codes was further streamlined by Lim, Feng, Pastore, Nazer, and Gastpar \cite{Lim--Gastpar2016}. With the primary motivation of communicating linear combinations of codewords over a multiple access channel (as in compute--forward \cite{Nazer--Gastpar2011,Hern--Narayanan2013}), they augmented
the original nested coset coding schemes in \cite{Padakandla--Pradhan2013, Miyake2010} by the channel transformation technique by Gallager \cite[Sec. 6.2]{Gallager1968} and developed new analysis tools when multiple senders use nested coset codes with a common generator matrix. The resulting achievable rate region, when adapted to the Gaussian case, improves upon the previous result for compute--forward \cite{Nazer--Gastpar2011}.

In both \cite{Padakandla--Pradhan2013} and \cite{Lim--Gastpar2016}, however, structured coding of nested coset codes is reserved for rather niche communication scenarios of adapting multiple codewords to a common channel state or computing sums of codewords, and even in these limited cases, as a complement to random coding. The coding scheme in \cite{Padakandla--Pradhan2013} uses superposition of codewords with individual and common generator matrices. A similar coding scheme in \cite{Padakandla--Pradhan2016} for three-user interference channels again uses a combination of random coding (for message decoding) and structured coding (for function decoding) of nested coset codes, this time with a more explicit superposition coding architecture.
There is also some indication that the benefit of computation can be realized to the full extend only in special cases for which desired linear combinations and channel structures are matched \cite{Karamchandani--Niesen--Diggavi2013}. In the same vein, the aforementioned rate region for computing in \cite{Lim--Gastpar2016} turns out to be strictly smaller than the typical capacity region, when computation is specialized to communication (i.e., the identity function computation). The authors of \cite{Lim--Gastpar2016} have recently improved their analysis to establish a larger achievable rate region for message communication~\cite{Lim--Gastpar2017}, which is still strictly smaller than the capacity region. Apparently, structured coding, even based on the promising new technique of nested coset codes, can only play a complementary role to random coding.

This paper aims to illustrate that at least for simple communication networks, the opposite is true, and that structured coding can completely replace random coding. In particular, we show that a random ensemble of nested coset codes of the same generator matrix (we referred to as \emph{homologous codes}~\cite{Sen--Kim2017}), which was thought to be good only for recovering linear combinations, can achieve the same rates as independently generated linear or nonlinear random codes for the task of communicating individual codewords over MACs. For simplicity of exposition, we start with two-sender MACs and show that the capacity region is achievable by a careful construction of random homologous codes. Our finding relies on the identification of \emph{shaping} and \emph{channel transformation} techniques, both of which are used to improve upon conventional coset codes by allowing nonuniform codewords, as key components to supplant random coding by structured coding. We first evaluate achievable rates of individual techniques, which fall short of the capacity region. We then combine these two techniques to obtain the best performance possible by \emph{any} transmission scheme. These results are extended to MACs with more than two senders, and with one or more receivers. Also, the achievability of the capacity region for the Gaussian counterparts is shown via an unconventional quantization argument that is consistent with the construction of homologous codes.

The rest of the paper is organized as follows. Section~\ref{sec:hnc_defn} defines nested coset codes and homologous codes. Section~\ref{sec:mot_exm} discusses the running examples of binary adder and binary erasure multiple access channels. The main results for two-sender MACs are presented in Section~\ref{sec:main_result}, and are extended to more than two senders and one or more receivers in Section~\ref{sec:k_user}. Section~\ref{sec:gauss} presents the achievability of the capacity region for Gaussian MACs. Section~\ref{sec:conc} concludes the paper by discussing the problem of simultaneous communication and computation, and the benefit of homologous coding.

We adapt the notation in~\cite{Cover--Thomas2006, El-Gamal--Kim2011}.
The set of integers $\{ 1,2,\ldots, n \}$ is denoted by $[1:n]$.
For a length-$n$ sequence (vector) $x^n=(x_1,x_2,\ldots,x_n) \in \Xc^n$, we define its type
as $\pi(x | x^n) = {|\{ i \suchthat x_i = x \}|}/{n}$ for $x \in \Xc$. Upper case letters $X,Y,\ldots$ denote random variables. 
For $\e \in (0,1)$, we define the $\e$-typical set of $n$-sequences (or the typical set in short) as $\aep(X) = \{ x^n \suchthat | p(x) - \pi(x| x^n) | \le \e p(x), \, x \in \Xc \}$. A tuple of $k$ random variables $(X_1, X_2, \ldots, X_k)$ is denoted by $X^k$, and for $\Jc \sbq [1:k]$, the subtuple of random variables with indices from $\Jc$ is denoted by $X_{\Jc} = (X_i: i \in \Jc)$. The indicator function $\mathbbm{1}_{\Sc}: \Xc \to \{ 0, 1 \}$ for $\Sc \sbq \Xc$ is
defined as $\mathbbm{1}_{\Sc}(x) = 1$ if $x \in \Sc$ and $0$ otherwise. A length-$n$ vector of all zeros (ones) is denoted by $\mathbf{0}_n$ ($\mathbf{1}_n$), where the subscript is omitted when it is clear in the context.
An  $m \times n$ matrix of all zeros is denoted by $O_{m \times n}$. The $n \times n$ identity matrix
is denoted by $I_n$.

\section{Homologous Codes}
\label{sec:hnc_defn}
A \emph{nested coset} code was first proposed in~\cite{Miyake2010}. 
Defined on a finite field $\Fq$ of order $q$, an $(n,k,\kh)$ nested coset code is defined by a $(k+\kh) \times n$ generator matrix $G$, a length-$n$ dithering vector (coset leader) $d^n$, and a shaping function $l: \Fq^k \to \Fq^{\kh}$. Let 
\begin{align}
\label{eqn:potential_code_ptp}
x^n(m,l) = [m \;\: l] \: G + d^n, \quad m \in \Fq^k, \; l \in \Fq^{\kh}.
\end{align}
Each message $m \in \Fq^{k}$ is then assigned a codeword $x^n(m,l(m))$, where $l(m)$ is the specified shaping function. A standard \emph{coset} code can be seen as a special case of a nested coset code with $\kh = 0$ (no shaping). Specializing further, we recover a \emph{linear code} as a nested coset code with $\kh = 0$ and $d^n = \Ob_n$.

Introduced in~\cite{Padakandla--Pradhan2013}, a \emph{random} nested coset code is an ensemble of nested coset codes that are constructed via a random generator matrix ${G}$ and a random dithering vector $D^n$ to emulate the behavior of a random (nonlinear) code ensemble drawn from a specified pmf $p(x)$ on $\Fq$. Each element of ${G}$ and $D^n$ is i.i.d.\@ $\U(\Fq)$. Given the realizations of ${G}$ and $D^n$, $x^n(m,l)$ for $m \in \Fq^k, \; l \in \Fq^{\kh}$ is defined as in (\ref{eqn:potential_code_ptp}). For shaping, we use the joint typicality encoding in~\cite{Padakandla--Pradhan2013}; see~\cite{Tal--Erez2008} for a similar technique in the context of lattice-based source coding. Let $p(x)$ be the desired pmf and $\e' > 0$. For each message $m \in \Fq^{k}$, choose an $l \in \Fq^{ \kh}$ such that $x^n(m,l) \in \aepvar(X).$ If there are more than one such $l$, choose one of them at random; if there is none, choose one in $\Fq^{\kh}$.

As shown in~\cite{Padakandla--Pradhan2013,Lim--Gastpar2016}, random nested coset code ensembles can achieve the capacity of a discrete memoryless channel $p(y | x)$. When the input alphabet $\Xc$ is not isomorphic to a finite field, the channel can be transformed into a virtual channel $p(y | u)$ with equal capacity via an appropriately chosen auxiliary input $U$ and symbol-by-symbol mapping $X = \vf (U)$. This result can be extended to the Gaussian channel~\cite{Lim--Gastpar2016} (via a quantization argument) and to multiple access channels~\cite{Padakandla--Pradhan2013}. In particular, for the $k$-sender discrete memoryless (DM) MAC $p(y | x_1,x_2, \ldots, x_k)$ and input pmfs $p(x_1), p(x_2), \ldots, p(x_k)$, each sender can use a random nested coset code ensemble (with individual generator matrices $G_1, G_2, \ldots, G_k$) to achieve the region $\Rrmac (X^k)$ characterized by 
\begin{align}
\sum_{i \in \Sc} R_i &< I(X_{\Sc};Y | X_{\Sc^c}), \quad \forall \Sc \sbq [1:k]. 
\label{eq:pentagon_mac}
\end{align}
Thus, heterologous nested coset codes (= with different generators) can emulate the performance of typically nonlinear random code ensembles for MACs. (In fact, for $k=2$, by controlling the structure of $G_1$ and $G_2$ more carefully, they can achieve larger rates than random codes for channels with state~\cite{Padakandla--Pradhan2013}).

We now consider nested coset codes with closer structural relationship. A collection of $(n,k_j,\kh_j)$ nested coset codes, $j=1,2,\ldots,N$, is said to be \emph{homologous} if they share a common generator matrix $G$ (but have individual dithering sequences and shaping functions). Since $k_j + \kh_j$ may differ, we use zero padding, i.e., instead of (\ref{eqn:potential_code_ptp}), we have 
\[ x_j^n(m_j,l_j) = [m_j \; l_j \; \Ob_{\kappa - (k_j + \kh_j)}] G + d_j^n,\]
where $\kappa = \max_j (k_j + \kh_j)$. In biological analogy, even though homologous codes are constructed from the same generator matrix, the actual ``shape" of the codes can be quite different due to individual shaping functions. \emph{Random} homologous codes are generated by a common generator matrix $G$ and dithering vectors $D_1^n, D_2^n, \ldots,D_N^n$ of i.i.d.\@ $\U(\Fq)$ entries, and shaping functions that find an $l_j$ such that \[ x_j^n(m_j,l_j) \in \aepvar(X_j), \quad j=1,2,\ldots,N,\] for given pmfs $p(x_1),p(x_2),\ldots,p(x_N)$.

Due to the use of a common generator matrix, homologous codes can achieve high rates when the goal of communication is to recover linear combination of codewords. For a $2$-sender DM-MAC, an achievable rate region is characterized in~\cite{Lim--Gastpar2016} for recovering linear combinations of codewords from random homologous code ensembles. When recovering both messages, however, this achievable rate region is in general smaller than the region in (\ref{eq:pentagon_mac}). Even a tighter probability of error analysis discussed in~\cite{Lim--Gastpar2017} does not guarantee the achievability of the region in (\ref{eq:pentagon_mac}). This raises the question of whether random homologous codes are useful only for communicating the sum of the messages (or the codewords) and fundamentally deficient compared to heterologous ones in communicating the messages themselves.

\section{Motivating Examples}
\label{sec:mot_exm}
We present two toy examples that illustrate the performance of homologous codes
and motivate our main result in Section~\ref{sec:main_result}. 
\begin{example}[Binary adder MAC]
\label{example:binary_add_mac}
Let $Y = X_1 \oplus X_2$, where $\Xc_1 = \Xc_2= \Yc = \{ 0, 1 \}$ 
and the addition operation $\oplus$ is over ${\GF(2)}$. The capacity region of this channel is achieved by
random coding with i.i.d.\@ $\Bern(1/2)$ inputs $X_1$ and $X_2$, and is depicted in
Fig.~\ref{fig:binary_add_mac_cap}. No binary linear or coset codes of the same generator matrix, however,
can achieve this region. As a matter of fact, binary linear or coset codes of the same generator matrix can only
achieve the rate region depicted in Fig.~\ref{fig:binary_add_mac_rates}. Achievability of this rate region is trivial. 
For the other direction, suppose without loss of generality that $R_1 \geq R_2>0$. 
Any message pair $({m}_1,{m}_2) \in \GF(2)^{nR_1} \times \GF(2)^{nR_2}$ 
results in the same output as the message pair 
$(m_1 \oplus  [{m} \; \Ob], m_2 \oplus m)$ for some $m \neq \Ob \in \GF(2)^{nR_2}$, which implies
the converse.

By using \emph{nested} coset codes with proper shaping, however, the capacity region can be achieved.
Suppose without loss of generality that $R_1 \ge R_2$ where $R_1+R_2 = 1$. Let $m_j \in \GF(2)^{nR_j}$ for $j=1,2$ and $l_2 \in \GF(2)^{nR_1}$, and consider
\begin{equation} \label{eq:nested_binary_adder}
\begin{split}
x_1^n(m_1) &= [m_1 \, ~\Ob], \\
x_2^n(m_2, l_2(m_2)) &= [m_2 \; l_2(m_2)] = [m_2 \, ~\Ob~ \, m_2].
\end{split}
\end{equation}
It is easy to see that this pair of homologous $(n, nR_1, 0)$ and $(n, nR_2, nR_1)$ codes with the same generator matrix $I_{n \times n}$ and trivial shaping function $l_2$
can communicate $m_1$ and $m_2$ without any error. 
\begin{figure}[t]
\begin{subfigure}[t]{0.5\textwidth}
\center
\includegraphics[scale=0.7]{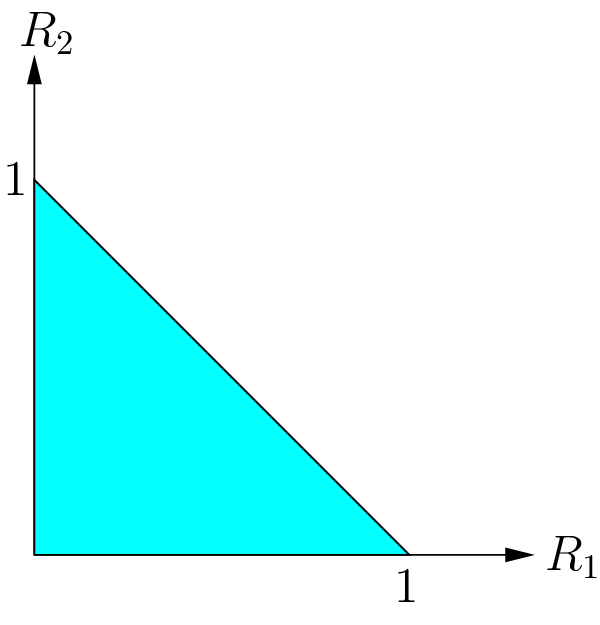}
\caption{}
\label{fig:binary_add_mac_cap}
\end{subfigure}%
\begin{subfigure}[t]{0.5\textwidth}
\center
\includegraphics[scale=0.7]{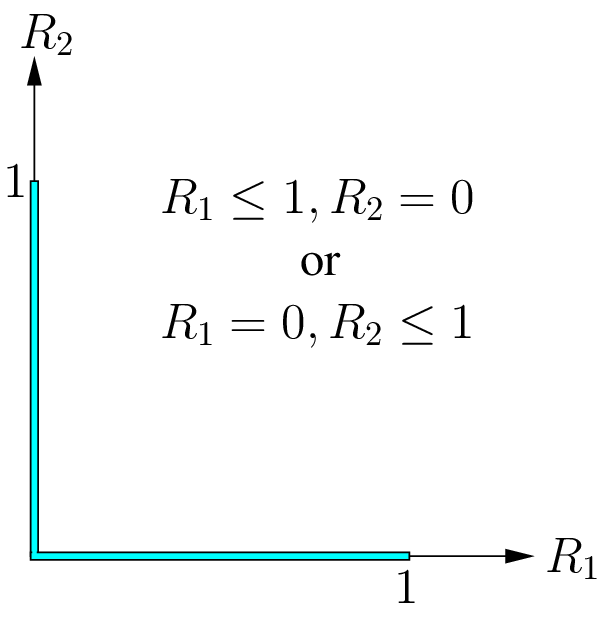}
\caption{}
\label{fig:binary_add_mac_rates}
\end{subfigure}
\caption{(a) The capacity region and (b) an achievable rate region of the binary adder MAC.}
\end{figure}
\end{example}

The next example has the underlying channel structure that is not
fully compatible with the algebraic structure of codes. 

\begin{example}[Binary erasure MAC]
\label{example:binary_erasure_mac}
Let $Y = X_1 + X_2$, where $\Xc_1 = \Xc_2= \{ 0, 1 \}$, $\Yc = \{ 0,1,2 \}$, and the addition operation $+$ is over $\Real$. The capacity region of the channel is
achieved by random coding with i.i.d.\@ $\Bern(1/2)$ inputs $X_1$ and $X_2$, and is depicted in
Fig.~\ref{fig:binary_erasure_hnc}. In contrast, no pair of binary coset codes with the same generator matrix can achieve the rate pair $(1/2 + \e, 1/2 + \e)$ for $\e > 0$. The proof of this claim is given in Appendix A.

This limitation of 
coset codes can be once again overcome by nested coset codes. Let $A_{k \times n}$ be a generator matrix of a linear code of rate $R = k/n < 1/2$ for the \emph{point-to-point} 
binary erasure channel of erasure probability $1/2$. Then, the following pair of linear codes
(with zero padding) can achieve the rate pair $(R, 1)$:
$x_1^n(m_1) = [m_1 \, \mathbf{0}_{n(1-R)}] \, B$ and
$x_2^n(m_2) = m_2 \, B$,
where 
\[ B = \begin{bmatrix}
A \\
A^\perp
\end{bmatrix},\] 
and $A^\perp$ is an $(n-k) \times n$ matrix whose rows are orthogonal to the rows of $A$.
We now construct homologous $(2n, n+k, 0)$ and $(2n, n+k, n-k)$ codes with the generator matrix
\begin{equation} \notag
G = \left[
\begin{array}{c|c}
B & O_{n \times n} \\
\hline
\begin{array}{c}
O_{k \times n}\\
A^\perp
\end{array}
& B
\end{array}
\right],
\end{equation}
and shaping function $l_2:GF(2)^{n+k} \rightarrow GF(2)^{n-k}$ such that $l_2(m_2) = \{m_{2i}\}_{i=k+1}^{n}$ for $m_2 \in GF(2)^{n+k}$. Then, it can be shown that the first and second halves of codewords are reliably communicated
at rates $(1,R)$ and $(R,1)$, which, combined together, can be arbitrarily close to $(3/4, 3/4)$. A similar argument can be extended to the entire capacity region.
\begin{figure}[t]
\begin{subfigure}[t]{0.5\textwidth}
\center
\includegraphics[scale=0.7]{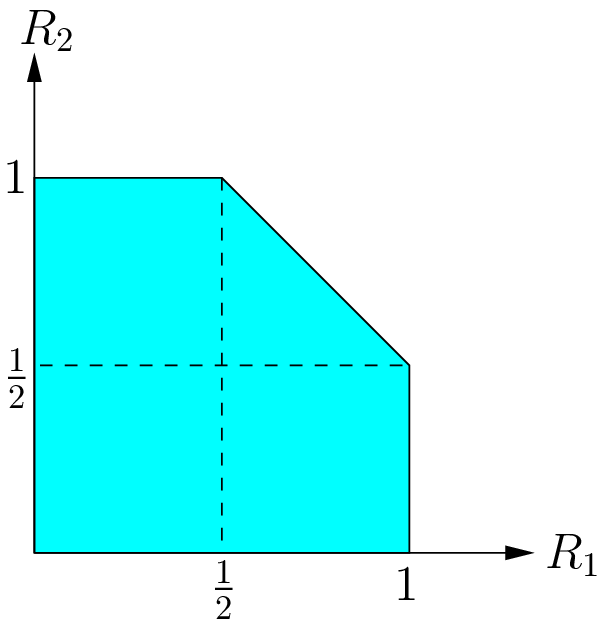}
\caption{}
\label{fig:binary_erasure_hnc}
\end{subfigure}%
\begin{subfigure}[t]{0.5\textwidth}
\center
\includegraphics[scale=0.7]{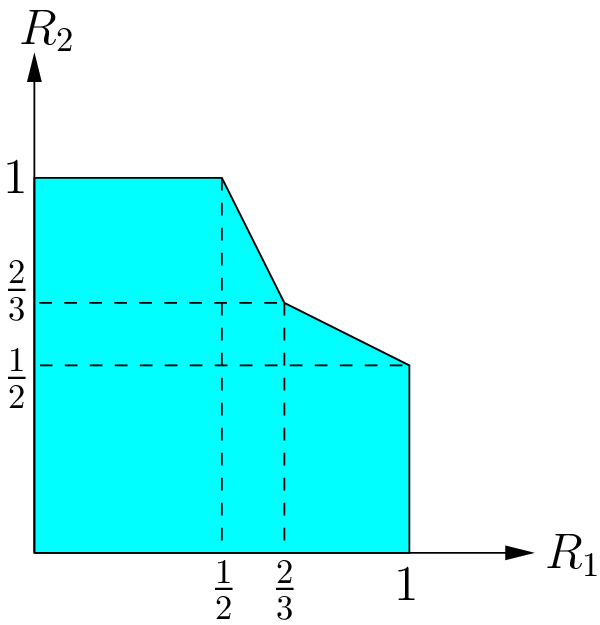}
\caption{}
\label{fig:shaping_transform}
\end{subfigure}
\caption{(a) The capacity region and (b) an achievable rate region of the binary erasure MAC.}
\end{figure}
\end{example}

The constructions of nested coset codes for binary adder and erasure MACs
respectively emulate time division and time sharing in disguise.
Consequently, these codes do not scale to more complicated problems (such as interference channels)
in a satisfactory manner. As we will illustrate shortly, however, 
most (random) homologous codes are sufficient to achieve the capacity region,
provided that they are constructed according to appropriate distributions.

\section{Achievable Rate Regions of Random Homologous Codes for Two Senders}
\label{sec:main_result}
We now investigate the performance of random homologous code ensembles defined
in Section~\ref{sec:hnc_defn}. Following the standard terminology in network information theory, we say that a rate pair $(R_1,R_2)$ is \emph{achievable} if there exists a sequence of codes of a fixed rate pair $(R_1,R_2)$ indexed by the block length $n$ such that the average probability of error $P_e^{(n)}$ tends to $0$ as $n\to \infty$. Specializing further, we say that $(R_1,R_2)$ is \emph{achievable by random homologous codes} if there exists a sequence of random homologous $(n,nR_1,n\Rh_1)$ and $(n,nR_2,n\Rh_2)$ code ensembles (cf. Section~\ref{sec:hnc_defn}) such that $\lim_{n \to \infty} \E [ P_e^{(n)} ] = 0$, where the expectation is taken with respect to the randomness in the common generator matrix and individual dithering sequences. 

We take a gradual approach to presenting the main result and first discuss the key technical ingredients of the proof one by one. Throughout, information measures are in log base $q$.

\subsection{Shaping}
\label{sec:shaping}
\begin{figure}[b]
\center
\includegraphics[scale=0.75]{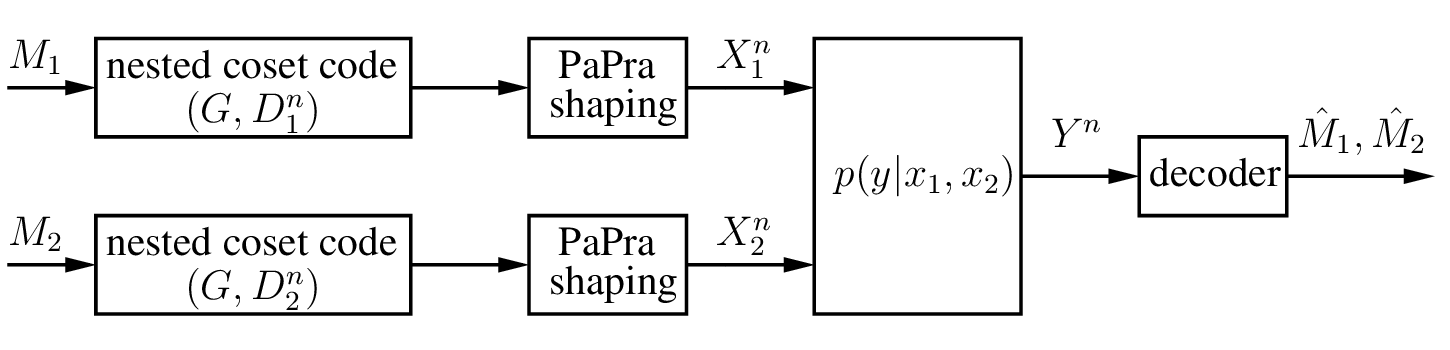}
\caption{Block diagram for the shaping technique.}
\label{fig:shaping_blk}
\end{figure}

Symbols in random coset codes are uniformly drawn over $\Fq$. By proper shaping via joint typically encoding, random homologous code ensembles emulate the statistical behavior of a random code ensemble while maintaining a common algebraic structure across users. The block diagram of this technique is depicted in Fig.~\ref{fig:shaping_blk}.

We describe the achievable rate region for a finite-field input DM-MAC $p(y | x_1,x_2)$, 
$\Xc_1 = \Xc_2 = \Fq$, by random homologous code ensembles. For given
input pmfs $p(x_1)$ and $p(x_2)$, we refer to the rate region in 
\eqref{eq:pentagon_mac} as $\Rrmac(X_1,X_2)$, i.e., the set of rate pairs $(R_1,R_2)$ such that
\begin{align*}
R_1 &< I(X_1; Y | X_2),\\
R_2 &< I(X_2; Y | X_1), \\
R_1 + R_2 &< I(X_1, X_2; Y),
\end{align*}
and define $\Rrl(X_1,X_2)$ as the set of rate pairs $(R_1,R_2)$ such that
\begin{equation} \label{eq:r1}
R_1 <  \max \{ I(X_1;Y),\, H(X_1)-H(X_2) +I(X_2;Y)  \}, 
\end{equation}
or
\begin{equation} \label{eq:r2}
R_2 <  \max \{ I(X_2;Y),\, H(X_2)-H(X_1) +I(X_1;Y)  \}.
\end{equation}

\begin{proposition} 
\label{prop:shaping_region}
A rate pair $(R_1,R_2)$ is achievable for the finite-field input DM-MAC $p(y | x_1,x_2)$ by random homologous codes if
\[
(R_1,R_2) \in \Rrmac(X_1,X_2) \cap \Rrl(X_1,X_2) 
\]
for some input pmfs $p(x_1)$ and $p(x_2)$.
\end{proposition}

\begin{IEEEproof}[Proof]
Our proof steps follow~\cite[Sec.~VI]{Lim--Gastpar2016} essentially line by line, except the analysis of one error event.
Fix $p(x_1)$ and $p(x_2)$. Let $\e' > 0$. We use random homologous code ensembles via typicality encoding (cf. Section~\ref{sec:hnc_defn}) constructed with the pmfs $p(x_1)$ and $p(x_2)$, and parameter $\e'$. The decoder finds a unique pair of $(\mh_1,\mh_2)$ such that
$(x_1^n(\mh_1,l_1), x_2^n(\mh_2,l_2),y^n) \in \aep(X_1,X_2,Y)$ for some $(l_1,l_2)$, where $\e > \e'$.
Assume that $(M_1, M_2)$ is the transmitted message pair and $(L_1, L_2)$ is the auxiliary index pair chosen
by the shaping functions. 
We bound the probability of error $\P(\Ec)$ averaged over codebooks.
As in~\cite{Lim--Gastpar2016}, the decoder makes an error only if one or more of the following events occur: 
\begin{align*}
\Ec_1 &= \{ X_j^n(M_j, l_j) \notin \aepvar (X_j) \text{ for all } l_j, \, j=1 \text{ or } 2 \}, 
\\
\Ec_2 &= \{  (X_1^n(M_1, L_1),X_2^n(M_2,L_2), Y^n) \notin  \aep (X_1,X_2,Y)\},
\\
\Ec_{3} &= \{  (X_1^n(M_1, L_1),X_2^n(m_2, l_2), Y^n) \in  \aep (X_1,X_2,Y) \text{ for some }   (m_2, l_2) \ne (M_2,L_2)  \},
\\
\Ec_{4} &= \{  (X_1^n(m_1, l_1),X_2^n(M_2, L_2), Y^n) \in  \aep (X_1,X_2,Y) \text{ for some}  (m_1, l_1) \ne (M_1,L_1)  \},
\\
\Ec_5 &= \{  (X_1^n(m_1, l_1),X_2^n(m_2, l_2), Y^n) \in  \aep (X_1,X_2,Y) \text{ for some }  (m_1, l_1) \neq (M_1,L_1), \\  & (m_2, l_2) \neq (M_2,L_2)  \text{ such that } [m_1 \; l_1 \; \Ob] \text{ and } [m_2 \; l_2 \; \Ob] \text{ are linearly independent}\},
\\
\Ec_6 &= \{  (X_1^n(m_1, l_1),X_2^n(m_2, l_2), Y^n) \in  \aep (X_1,X_2,Y) \text{ for some }  (m_1, l_1) \neq (M_1,L_1),\\
& (m_2, l_2) \neq (M_2,L_2)  \text{ such that } [m_1 \; l_1 \; \Ob] \text{ and } [m_2 \; l_2 \; \Ob] \text{ are linearly dependent}\}.
\end{align*}%
Thus, by the union of evens bound, $\P(\Ec) \le \P(\Ec_1) + \sum_{k\ne 1} \P(\Ec_k \cap \Ec_1^c)$. 
By~\cite{Lim--Gastpar2016}, the first five terms tend to $0$ as $n \to \infty$ if 
\begin{align} \notag
\Rh_j & > D_j + \d(\e'), \; j=1,2
\\ \notag
R_1 + 2 \Rh_1 + \Rh_2 & < I(X_1;Y| X_2)  + 2D_1 + D_2 - \d(\e),
\label{reg:lim_constraints}\\
R_2 +  \Rh_1 + 2 \Rh_2 & < I(X_2;Y| X_1)  + D_1 + 2 D_2 - \d(\e),
\\ \notag
R_1 + R_2 + 2 \Rh_1 + 2 \Rh_2 & < I(X_1,X_2;Y)  + 2D_1 + 2D_2 - \d(\e),
\end{align}
where $D_j \triangleq D(p_{X_j} | | \U(\Fq))$, $j=1,2$. 
For the last term, the authors of \cite{Lim--Gastpar2016} provide an upper bound on $R_1$ and $R_2$ in terms of two linear combinations of $U_1$ and $U_2$, namely, $W_1 = a_1 U_1 + a_2 U_2$ and $W_2 = b_1 U_1 + b_2 U_2$ that are linearly independent. We present a new upper bound, which results in a larger achievable rate region than substituting $(W_1,W_2) = (X_1,X_2)$ or $(W_1,W_2) = (X_2,X_1)$ in the rate region provided by \cite{Lim--Gastpar2016}. 

\begin{lemma}
\label{lem:proof_of_last_term}
The probability $\P(\Ec_6 \cap \Ec_1^c)$ can be bounded by two different expressions: 
\begin{align*}
\P(\Ec_6 \cap \Ec_1^c) & \le  (q-1) q^{n(\Rh_1+\Rh_2+\min \{{ {R}_1+\Rh_1}, {R}_2+\Rh_2 \})} q^{n(H(X_1)+H(X_2)+H(X_2 | Y)-3+\d(\e))}, \\
\P(\Ec_6 \cap \Ec_1^c) & \le (q-1)q^{n(\Rh_1+\Rh_2+\min \{{ {R}_1+\Rh_1}, {R}_2+\Rh_2\})} q^{n(H(X_1)+H(X_2)+H(X_1 | Y)-3+\d(\e))}.
\end{align*}
\end{lemma}

\begin{IEEEproof}
Define the rate $R = \min \{ R_1 + \Rh_1, R_2 + \Rh_2 \}$, and
the events $\Mc = \{M_1 = \Ob,M_2 = \Ob \}$ and
$\Lc = \{L_1 = \Ob,L_2 = \Ob \}$. Define the set
\[
\Dc = \{(m_1,l_1,m_2,l_2) \in \Fq^{nR_1} \times \Fq^{n \Rh_1} \times \Fq^{nR_2} \times \Fq^{n\Rh_2}:
[m_1 \; l_1 \; \Ob] \neq \Ob,
[m_2 \; l_2 \; \Ob] \neq \Ob \text{ are linearly dependent} \}.
\]
By the symmetry of code generation, $\P(\Ec_6 \cap \Ec_1^c) = \P(\Ec_6 \cap \Ec_1^c | \Mc,\Lc)$, which is bounded by
\begin{align*}
& \P  (\Ec_6  \cap \Ec_1^c  | \Mc,\Lc) = \P(X_1^n(m_1, l_1),X_2^n(m_2, l_2), Y^n) \in  \aep (X_1,X_2,Y) \\
& \quad\quad\quad\quad\quad\quad\quad\quad\quad\quad\quad\quad
\text{for some }  (m_1, l_1, m_2, l_2) \in \Dc, X_j^n(\Ob, \Ob) \in \aepvar (X_j) \; j=1,2 | \Mc, \Lc)
\\
& \overset{(a)}{\le} \sum\limits_{(m_1, l_1, m_2, l_2) \atop \in \Dc} 
\P(X_1^n(m_1, l_1),X_2^n(m_2, l_2), Y^n) \in  \aep (X_1,X_2,Y),
 X_j^n(\Ob, \Ob) \in \aepvar (X_j) \; j=1,2 | \Mc, \Lc)
\\
& \le \sum\limits_{(m_1, l_1, m_2, l_2) \in \Dc} 
\P(X_2^n(m_2, l_2), Y^n) \in  \aep (X_2,Y),
 X_j^n(\Ob, \Ob) \in \aepvar (X_j) \, j=1,2 | \Mc, \Lc)
\\
& \le \sum\limits_{(m_1, l_1, m_2, l_2) \in \Dc}  \,
\sum_{x_1^n \in \aep(X_1), \atop x_2^n \in \aep(X_2)}
\sum_{(\xt_2^n,y^n) \in \atop   \aep (X_2,Y)} 
\P\left(  \left. \begin{array}{c}
[m_2 \; l_2 \; \Ob] {G} + D_2^n = \xt_2^n, \\
Y^n = y^n, D_1^n = x_1^n, D_2^n = x_2^n
\end{array}  \right \rvert \Mc, \Lc \right) 
\\
& \overset{(b)}{=} \sum\limits_{(m_1, l_1, m_2, l_2) \in \Dc} \,
\sum_{x_1^n \in \aep(X_1), \atop x_2^n \in \aep(X_2)}
\sum_{(\xt_2^n,y^n) \in \atop   \aep (X_2,Y)} 
\P\left(  \left.  \begin{array}{c}
[m_2 \; l_2 \; \Ob]{G} + D_2^n = \xt_2^n, \\
D_1^n = x_1^n, D_2^n = x_2^n
\end{array}  \right \rvert \Mc, \Lc \right) p(y^n|x_1^n,x_2^n)
\\
& \overset{(c)}{\le} q^{n(\Rh_1 + \Rh_2)}  
\sum\limits_{(m_1, l_1, m_2, l_2) \in \Dc} \,  
\sum_{x_1^n \in \aep(X_1), \atop x_2^n \in \aep(X_2)}
\sum_{y^n \in \atop   \aep (Y)}  p(y^n \vert x_1^n, x_2^n) 
\sum_{\xt_2^n \in \atop \aep (X_2 \vert y^n)}  
\P\left(  \begin{array}{c}
[m_2 \; l_2 \; \Ob]{G} + D_2^n = \xt_2^n, \\
D_1^n = x_1^n, D_2^n = x_2^n
\end{array}  \right) 
\\
& \le q^{n(\Rh_1 + \Rh_2)}  
\sum\limits_{(m_1, l_1, m_2, l_2) \in \Dc} \, 
\sum_{x_1^n \in \aep(X_1), \atop x_2^n \in \aep(X_2)}
\sum_{y^n \in \atop   \aep (Y)}  p(y^n \vert x_1^n, x_2^n) 
\sum_{\xt_2^n \in \atop \aep (X_2 \vert y^n)}  
q^{-3n}
\\
& \le  q^{n(\Rh_1 + \Rh_2)} \,  |\Dc| \, q^{n(H(X_1) + H(X_2)+ H(X_2\vert Y) + \delta(\epsilon))}  q^{-3n},
\\
& \leq q^{n(\hat{R}_1 + \hat{R}_2)} (q-1) q^{nR} q^{n(H(X_1) + H(X_2)+ H(X_2\vert Y) + \delta(\epsilon))}  q^{-3n},
\end{align*}
where $(a)$ follows by the union of events bound, $(b)$ follows since, conditioned on $(\mathcal{M},\mathcal{L})$, the triple ${G} \rightarrow (D_1^n,D_2^n) \rightarrow Y^n$ form a Markov chain,
and $(c)$ follows by \cite[Lemma~11]{Lim--Gastpar2016}. By changing the order of $X_1^n$ and $X_2^n$, we obtain the second bound on $\P(\mathcal{E}_6 \cap \mathcal{E}_1^c)$.     
\end{IEEEproof}

By Lemma~\ref{lem:proof_of_last_term}, $\P(\mathcal{E}_6 \cap \mathcal{E}_1^c)\rightarrow 0$ as $n \rightarrow \infty$ if $\min \lbrace R_1 +  2\hat{R}_1 + \hat{R}_2, R_2 + \hat{R}_1 + 2 \hat{R}_2 \rbrace < H(X_1) + 2D_1 \allowbreak + \allowbreak D_2 \allowbreak - \allowbreak \min \lbrace H(X_1 \vert Y), H(X_2 \vert Y) \rbrace - \delta(\epsilon)$. Choosing $\hat{R}_1 = D_1 + 2\delta(\epsilon'),\hat{R}_2 = D_2 + 2\delta(\epsilon')$ and letting $\epsilon \rightarrow 0$ yield the achievable rate region that consists of the rate pairs $(R_1,R_2)$ such that
\begin{align}
\label{reg:hnc_finite} \notag
R_1 &< I(X_1;Y | X_2), \\ \notag
R_2 &< I(X_2;Y | X_1), \\
R_1 + R_2 &< I(X_1,X_2;Y), \\ \notag
\min \{ R_1  +  H(X_2), R_2  +  H(X_1) \} &< 
H(X_1) + H(X_2)-\min \{ H(X_1 | Y), H(X_2 | Y) \}.
\end{align}
The rate region defined by (\ref{reg:hnc_finite}) is equivalent to the region $\Rrmac(X_1,X_2) \cap \Rrl(X_1,X_2)$, as will be proved in Appendix B.
\end{IEEEproof}

When specialized to the binary adder MAC, the achievable rate region in Proposition~\ref{prop:shaping_region} is indeed equivalent to the capacity region, which is proved in Appendix C. For the binary erasure MAC, however, the rate region in Proposition~\ref{prop:shaping_region} is \emph{strictly smaller} than the capacity region, as sketched in Fig.~\ref{fig:shaping_transform}.
In particular, the largest achievable symmetric rate is $2/3$ (see Appendix D).

We now introduce another simple example, which will be used again in Section~\ref{sec:compound} when we deal with multiple-receiver MACs.

\begin{example}[On--off erasure MAC]
\label{example:onoff_erasure_mac}
Let $Y = (2X_1-1) + Z (2X_2-1)$, where $\Xc_1 = \Xc_2 = \{0,1\}$ and
$\Yc = \{ 0, \pm 1, \pm 2 \}$, and random variable $Z \sim \Bern(p)$ is independent from $X_1$ and $X_2$. If $Z=1$, the channel behaves very similarly to the binary erasure MAC. If $Z=0$, the output $Y$ is only dependent on $X_1$. That is why this channel is called as the \emph{on--off erasure MAC}. 

For an arbitrary $p>0$, the capacity region of the on--off erasure MAC is achieved by random coding with i.i.d.\@ $\Bern(1/2)$ inputs $X_1$ and $X_2$, and is shown in Fig.~\ref{fig:onoff_mac_cap} (in terms of $p$). The achievable rate region in Proposition~\ref{prop:shaping_region} is equivalent to the capacity region for the on--off erasure MAC with $p \le 2/3$. For $p > 2/3$, however, it reduces to the rate region depicted in~\ref{fig:onoff_mac_shaping} that is strictly smaller than the capacity region (see Appendix~E). Note that for $p=1$, the rate region in~\ref{fig:onoff_mac_shaping} is equivalent to the achievable rate region for the binary erasure MAC sketched in Fig~\ref{fig:shaping_transform},
since the on--off erasure MAC behaves exactly as the binary erasure MAC when $p=1$.

\begin{figure}[tb]
\begin{subfigure}[t]{0.5\textwidth}
\center
\includegraphics[scale=0.75]{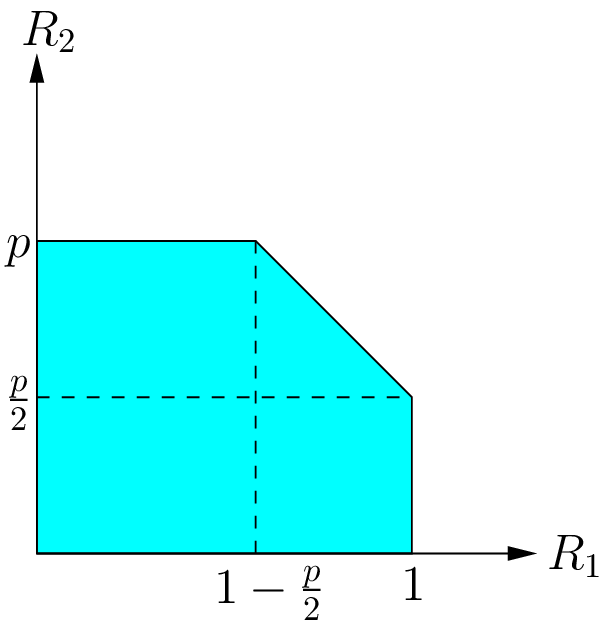}
\caption{}
\label{fig:onoff_mac_cap}
\end{subfigure}%
\begin{subfigure}[t]{0.5\textwidth}
\center
\includegraphics[scale=0.75]{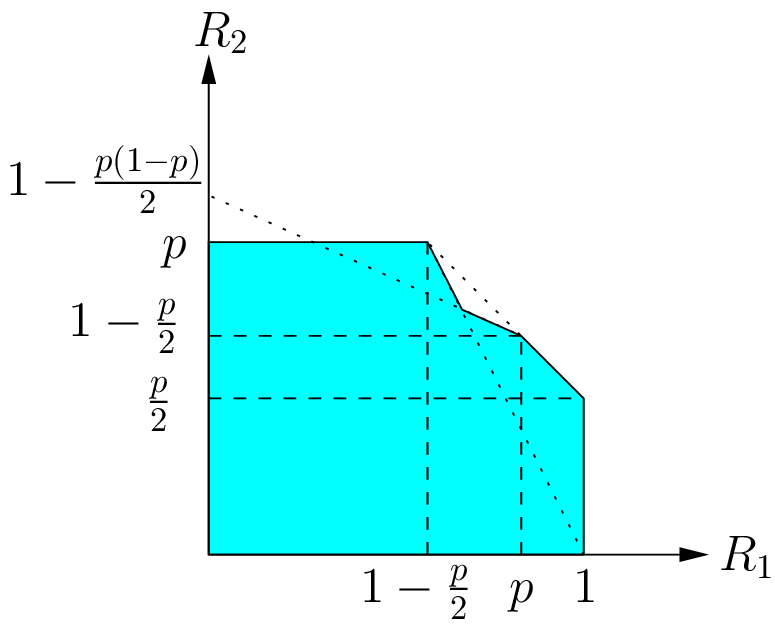}
\caption{}
\label{fig:onoff_mac_shaping}
\end{subfigure}
\caption{(a) The capacity region and (b) an achievable rate region of the on--off erasure MAC for $p > 2/3$.}
\end{figure}
\end{example}

\begin{remark}
\label{rem:shaping}
As shown by \cite{Lim--Gastpar2017}, the achievable rate region in Proposition~\ref{prop:shaping_region} can be improved by stronger analysis tools, which we will discuss later in Section \ref{sec:shaping_k} and Proposition \ref{prop:shaping_region_k}. For Examples \ref{example:binary_add_mac}--\ref{example:onoff_erasure_mac}, however, the achievable rate region in \cite{Lim--Gastpar2017} reduces to that of Proposition~\ref{prop:shaping_region}. 
\end{remark}

\subsection{Channel Transformation}
\label{sec:channel_trans}
\begin{figure}[htb]
\center
\includegraphics[scale=0.8]{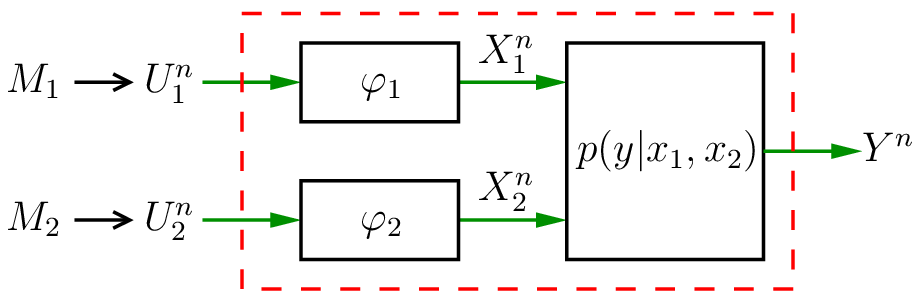}
\caption{The transformed DM-MAC $p(y | u_1, u_2)$ with virtual inputs $U_1$ and $U_2$.}
\label{fig:virtual_mac}
\end{figure}

Instead of using a nested coset code and choosing an appropriate shaping function,
there is a simpler way of achieving the performance of nonuniformly distributed codes.
Following the basic idea in~\cite{Gallager1968},
we can simply transform the channel $p(y | x_1,x_2)$ into a \emph{virtual channel} with finite-field inputs
\begin{equation} \label{eq:virtual_channel}
p(y | u_1,u_2) = p_{Y | X_1,X_2}(y | \vf_1(u_1),\vf_2(u_2))
\end{equation} 
for some symbol-by-symbol mappings $\vf_1: \Fq \to \Xc_1$ and $\vf_2: \Fq \to \Xc_2$, as illustrated in Fig.~\ref{fig:virtual_mac}. Note that this transformation can be applied to any DM-MAC $p(y|x_1,x_2)$ of arbitrary (not necessarily the same finite-field) input alphabets.

\begin{figure}[tb]
\center
\includegraphics[scale=0.75]{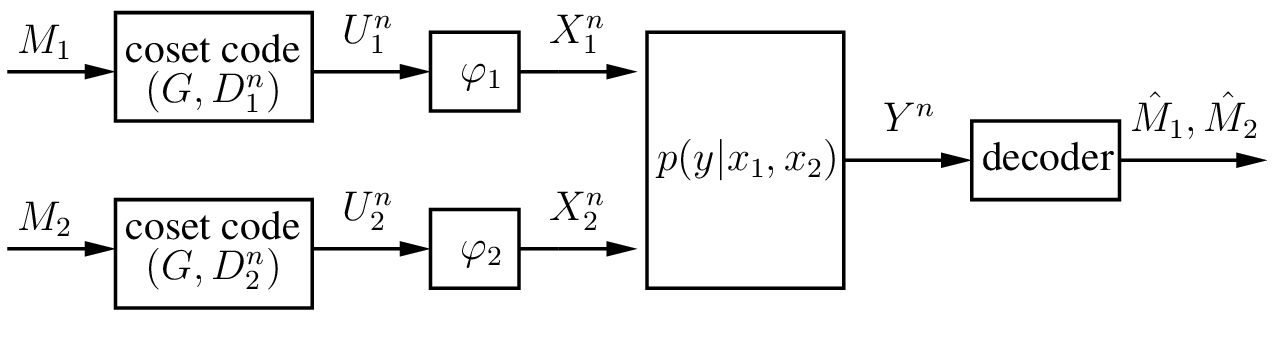}
\caption{Block diagram for the channel transformation technique.}
\label{fig:ct_blk}
\end{figure}

We now consider a pair of random coset codes of the same generator matrix for the virtual channel,
which is equivalent to random homologous codes with $\Rh_1 = \Rh_2 = 0$. The block diagram of this technique is depicted in Fig.~\ref{fig:ct_blk}.
Following the similar (yet simpler) steps to the proof of
achievability in Proposition~\ref{prop:shaping_region},
we can establish the following.

\begin{proposition}
\label{prop:chan_trans_prop_u}
A rate pair $(R_1,R_2)$ is achievable for the DM-MAC $p(y | x_1,x_2)$ by random coset codes 
in $\Fq$ with the same generator matrix, if 
\[
(R_1,R_2) \in  \Rrmac (U_1,U_2) \cap \Rrl (U_1,U_2),
\]
where $\Rrmac (U_1,U_2)$ is defined as in \eqref{eq:pentagon_mac}
for the virtual channel $p(y|u_1,u_2)$ in \eqref{eq:virtual_channel} and for the inputs $U_1$ and $U_2$ drawn independently according to $\U(\Fq)$,
and $\Rrl (U_1, U_2)$ is the set of $(R_1, R_2)$ such that 
\begin{equation} \label{eq:L}
\min(R_1,R_2) < \max\{ I(U_1; Y), \, I(U_2; Y)\}.
\end{equation}
\end{proposition}

Note that \eqref{eq:L} is equivalent to \eqref{eq:r1} and \eqref{eq:r2} with $(U_1,U_2)$ in place of $(X_1,X_2)$ since $U_1$ and $U_2$ are uniform on $\Fq$. The same region can be achieved by random \emph{linear} codes ($D_1^n = D_2^n = \mathbf{0}^n$) as well.

Proposition~\ref{prop:chan_trans_prop_u} was stated for a fixed channel transformation 
specified by a given pair of symbol-by-symbol mappings $\vf_1(u_1)$ and $\vf_2(u_2)$ on
a finite field $\Fq$.
We now consider all such channel transformations, which results in a simpler achievable rate region.

\begin{corollary}
\label{cor:phy_mac_coset}
A rate pair $(R_1,R_2)$ is achievable for the DM-MAC $p(y | x_1, x_2)$ 
by random coset codes in some finite field with the same generator matrix,
if 
\[
(R_1,R_2) \in \Rrmac (X_1,X_2) \cap \Rrl'(X_1,X_2)
\]
for some input pmfs $p(x_1)$ and $p(x_2)$, where
$\Rrl'(X_1, X_2)$ is the set of $(R_1, R_2)$ such that 
\[
\min(R_1,R_2) < \max\{ I(X_1; Y), \, I(X_2; Y)\}.
\]
\end{corollary}

\begin{IEEEproof}
First suppose that $p(x_1)$ and $p(x_2)$ are of the form 
\begin{equation} \label{eq:prime_pmf}
\frac{i}{p^m}
\end{equation}
for some prime $p$ and $i, m \in \Zz^+$ for all $x_1$ and $x_2$.
Then there exist $\vf_1(u_1)$ and $\vf_2(u_2)$ on $\Fq$
such that $X_j \overset{d}{=} \vf_j(U_j)$ with $U_j \sim \U(\Fq)$, where $q = p^m$.
Hence, we can transform the channel $p(y|x_1,x_2)$ into a virtual channel $p(y|u_1,u_2)$
and achieve the rate region in Proposition~\ref{prop:chan_trans_prop_u}. Now, since 
$(U_1,U_2) \to (X_1,X_2) \to Y$ form a Markov chain and $(U_1,X_1)$ and $(U_2,X_2)$ are independent, 
$\Rrl(U_1, U_2)$ in Proposition~\ref{prop:chan_trans_prop_u} can be simplified as
$\Rrl'(X_1, X_2)$. Finally, the earlier restrictions on the input pmfs can be removed
since the set of pmfs of the form \eqref{eq:prime_pmf} is dense. This completes the proof.
\end{IEEEproof}

For the binary adder MAC, the achievable rate region in Corollary~\ref{cor:phy_mac_coset}
is equivalent to the capacity region. To see this, note that for the binary adder MAC, $\Rrl(X_1,X_2) \sbq \Rrl'(X_1,X_2)$ for any $p(x_1)$ and $p(x_2)$, and the former region achieved by shaping (with the intersection with $\Rrmac(X_1,X_2)$) reduces to the capacity region as proved in Appendix C. Therefore, the capacity region of the binary adder MAC is achievable by using coset codes over the transformed channel. This does not contradict the fact that no coset code on the \emph{binary field} can achieve a positive symmetric rate pair, since channel transformation allows the use of linear (or coset) codes over larger finite fields.

For the binary erasure MAC, the channel transformation technique achieves the same rate region in Fig.~\ref{fig:shaping_transform} as the shaping technique (Proposition~\ref{prop:shaping_region}),
although $\Rrl'(X_1,X_2)$ is in general different than $\Rrl(X_1,X_2)$ for fixed pmfs $p(x_1)$ and $p(x_2)$. The proof is given in Appendix~D. 

\begin{figure}[tb]
\begin{subfigure}[t]{0.5\textwidth}
\center
\includegraphics[scale=0.75]{figures/onoff_mac_cap}
\caption{}
\label{fig:onoff_mac_cap2}
\end{subfigure}%
\begin{subfigure}[t]{0.5\textwidth}
\center
\includegraphics[scale=0.75]{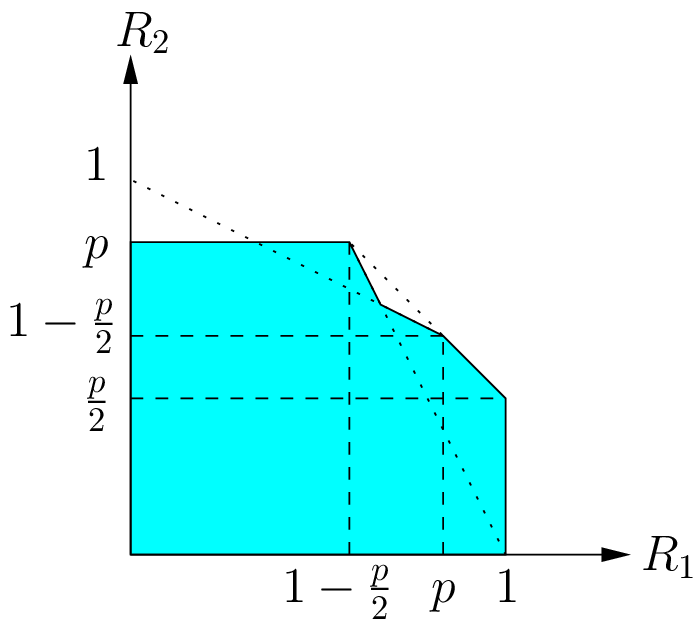}
\caption{}
\label{fig:onoff_mac_ct}
\end{subfigure}
\caption{(a) The capacity region and (b) an achievable rate region of the on--off erasure MAC for $p > 2/3$.}
\end{figure}

For the on--off erasure MAC with $p \le 2/3$, channel transformation achieves the capacity region sketched in Fig.~\ref{fig:onoff_mac_cap2}. For $p > 2/3$, however, it achieves the rate region sketched in Fig. \ref{fig:onoff_mac_ct}. While larger than what is achieved by shaping (cf.~Fig.~\ref{fig:onoff_mac_shaping}), the achievable rate region by channel transformation 
in Corollary~\ref{cor:phy_mac_coset}
is still strictly smaller than the capacity region. The details are given in Appendix E.

\begin{remark}
\label{rem:CT}
The achievable rate region for the channel transformation technique in Corollary \ref{cor:phy_mac_coset} can be easily evaluated for fixed input pmfs $p(x_1)$ and $p(x_2)$. Using the analysis tools developed in \cite{Lim--Gastpar2017}, Proposition \ref{prop:chan_trans_prop_u} and Corollary \ref{cor:phy_mac_coset} can be potentially strengthened. The resulting achievable rate region, however, is not computable. Therefore, it is unclear whether the insufficiency of the channel transformation technique for Examples \ref{example:binary_erasure_mac}--\ref{example:onoff_erasure_mac} (binary erasure MAC and on--of erasure MAC) is fundamental or due to the deficiency of our analysis tools. (We are unable to evaluate the larger region, which could be even loose).
\end{remark}

\subsection{Combination}
\begin{figure}[tb]
\center
\includegraphics[scale=0.75]{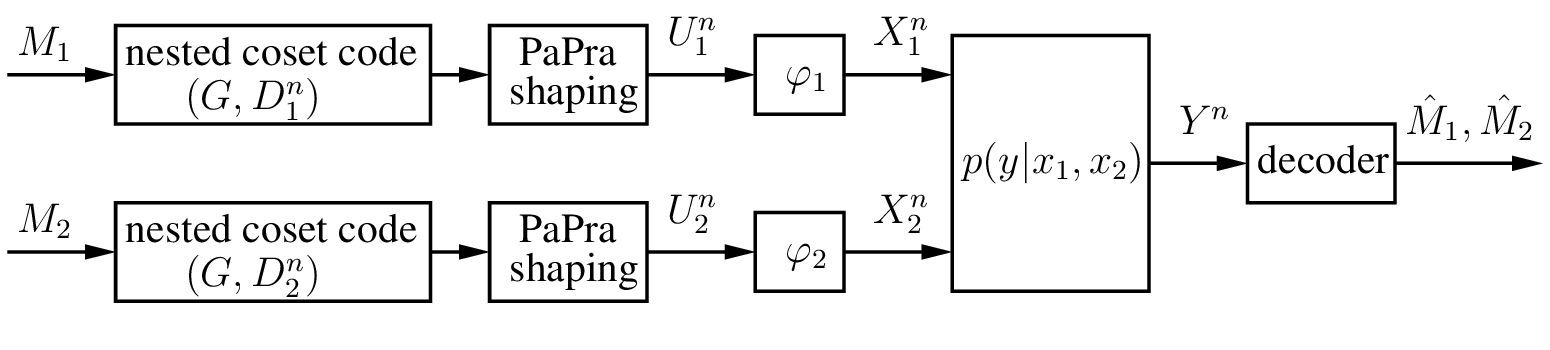}
\caption{Block diagram for homologous codes over the transformed channel.}
\label{fig:homologous_blk}
\end{figure}

As shown for the binary erasure MAC and on--off erasure MAC examples,
shaping (with homologous codes) and channel transformation
(with coset codes of the same generator matrix) seemingly cannot achieve the capacity
region. When combined together, these techniques can achieve the pentagonal region $\Rrmac(X_1,X_2)$ for
any $p(x_1)$ and $p(x_2)$ while maintaining the algebraic structure of the code. Consider the virtual channel in \eqref{eq:virtual_channel}
and random homologous codes for this channel, a block diagram for which is depicted in Fig.~\ref{fig:homologous_blk}. Then, Proposition~\ref{prop:shaping_region} implies
the following.
\begin{proposition}
\label{pro:phy_mac}
A rate pair $(R_1,R_2)$ is achievable for the DM-MAC $p(y | x_1,x_2)$
by random homologous codes in $\Fq$,
if
\[
(R_1,R_2) \in \Rrmac(X_1,X_2) \cap \Rrl(U_1,U_2,X_1,X_2)
\]
for some pmfs $p(u_1)$ and $p(u_2)$ on $\Fq$, and some mappings 
$x_1 = \vf_1(u_1)$ and $x_2 = \vf_2(u_2)$,
where $\Rrl(U_1,U_2,X_1, X_2)$ is the set of rate pairs $(R_1,R_2)$ such that
\begin{equation}
\label{eq:rl1}
R_1 <  \max \left[ I(X_1;Y), H(U_1)-H(U_2) +I(X_2;Y) \right]
\end{equation}
or
\begin{equation}
R_2 <  \max \left[ H(U_2)-H(U_1) + I(X_1; Y), I(X_2;Y)  \right].
\label{eq:rl2}
\end{equation}
\end{proposition}

We are now ready to state one of the main technical results of this paper, which follows from
Proposition~\ref{pro:phy_mac} by optimizing over all channel transformations.

\begin{theorem}
\label{thm:shaping_cap}
A rate pair $(R_1,R_2)$ is achievable for the DM-MAC $p(y | x_1,x_2)$
by random homologous codes in some finite field, if 
$
(R_1,R_2) \in \Rrmac(X_1,X_2)
$
for some $p(x_1)$ and $p(x_2)$.
\end{theorem}

\begin{IEEEproof}
Our argument is similar to the proof of Corollary~\ref{cor:phy_mac_coset}, except that the choice of 
channel transformation needs more care.
First suppose that $p(x_1)$ and $p(x_2)$ are of the form \eqref{eq:prime_pmf}. We will show that there exist a finite field $\Fq$, pmfs $p(u_1)$ and $p(u_2)$ on $\Fq$, and mappings $x_1 = \vf_1(u_1)$ and $x_2 = \vf_2(u_2)$ such that $\Rrmac(X_1,X_2) \sbq \Rrl(U_1,U_2,X_1,X_2)$. Consider random homologous codes over $\Fq$ with $q = p^{2m}$.
Choose $U_1$ and $\vf_1$ such that $U_1$ and $\vf_1(U_1) \overset{d}{=} X_1$
are one-to-one on the support of $U_1$ (this is always possible since $q \ge p^m$).
Also choose $U_2 \sim \U(\Fq)$ and $\vf_2$ such that $\vf_2(U_2) \overset{d}{=} X_2$
(this is possible due to the form of $p(x_2)$).
Let $(R_1,R_2) \in \Rrmac(X_1,X_2)$. Then, $(R_1,R_2)$ satisfies
\begin{align*}
R_2 &< I(X_2;Y|X_1) \\
& \le H(X_2) \\
& \le \log p^m \\
& \le H(U_2) - H(U_1) \\
& \le H(U_2) - H(U_1) + I(X_1;Y),
\end{align*}
which implies that $(R_1,R_2) \in \Rrl(U_1,U_2,X_1,X_2)$.
Finally, the restrictions on the input pmfs can be removed again by the denseness argument.
\end{IEEEproof}

\section{Extension to More Than Two Senders}
\label{sec:k_user}
The achievable rate region by random homologous codes for the $2$-sender DM-MAC can be extended to DM-MACs with more senders. Defining achievability of the rate tuples in a similar manner to the $2$-sender case, we present the performance of random homologous code ensembles for the $k$-sender DM-MAC $p(y|x_1,x_2,\ldots,x_k)$. Similar to Section~\ref{sec:main_result}, we first discuss the performance of random homologous codes under the fixed channel alphabets, following the recent work in~\cite{Lim--Gastpar2017}. We then generalize the result by incorporating channel transformation.

\subsection{Shaping}
\label{sec:shaping_k}
The achievable rate region for the finite-field input DM-MAC $p(y | x_1,x_2, \ldots,x_k)$, 
$\Xc_1 = \Xc_2 = \cdots= \Xc_k = \Fq$, by random homologous code ensembles was studied in~\cite{Lim--Gastpar2017}. For the sake of completeness, we review the main result in~\cite{Lim--Gastpar2017} on which we build the achievability of the capacity region for the $k$-sender DM-MAC. Let $\Ac$ denote the set of rank deficient $k \times k$ matrices over $\Fq$. For a given matrix $A \in \Ac$, we define the collection  
\[
\Dr(A) = \{ \Sc \sbq [1:k]: |\Sc| = k-\rank(A), \; 
\rank [A^T \; e(\Sc)^T]^T = k \},
\]
where $e(\Sc) \in \Fq^{|\Sc| \times k}$ denotes the matrix whose rows are the standard basis vectors $e_j$ for $j \in \Sc$.
For a given set $\Sc \in \Dr(A)$ and input pmfs $p(x_1),p(x_2),\ldots,p(x_k)$, we define the rate region $\Rr(A, \Sc, X^k)$ as the set of rate tuples
$(R_1,R_2,\ldots,R_k)$ such that
\begin{equation*} 
\sum_{j \in \Sc} R_j <  I(X_{\Sc};Y, W_A), 
\end{equation*}
where
\[
W_{A} = A \: [X_1 \; X_2 \; \ldots \; X_k]^T.
\]
We are now ready to state the main result of~\cite{Lim--Gastpar2017}.

\begin{proposition}[{\cite[Theorem 1]{Lim--Gastpar2017}}]
\label{prop:shaping_region_k}
A rate tuple $(R_1,R_2,\ldots,R_k)$ is achievable for the finite-field input DM-MAC $p(y | x_1,x_2,\ldots,x_k)$ by random homologous codes if
\[
(R_1,R_2,\ldots,R_k) \in \bigcap_{A \in \Ac} \; \bigcup_{\Sc \in \Dr(A)} \Rr(A, \Sc,X^k) 
\]
for some input pmfs $p(x_1), p(x_2),\ldots,p(x_k)$.
\end{proposition}

\begin{remark}[Revisit of the $2$-sender DM-MAC]
Consider the 2-sender DM-MAC $p(y|x_1,x_2)$ with given input pmfs $p(x_1)$ and $p(x_2)$. To compute the achievable rate region in Proposition~\ref{prop:shaping_region_k}, it suffices to consider the set of rank deficient $2 \times 2$ matrices with different spans. There are four types of such matrices: 
\begin{itemize}
\item $A = \left[\begin{array}{cc}
0 & 0 \\
0 & 0
\end{array} \right]$: $\Dr(A) = \{ \{ 1,2 \} \}$ and $\Rr(A, \{1,2\}, X_1,X_2)$ reduces to the set of rate pairs satisfying
\[
R_1 + R_2 < I(X_1,X_2;Y),
\]
\item $A = \left[\begin{array}{cc}
0 & 1 \\
0 & 0
\end{array} \right]$: $\Dr(A) = \{ \{ 1 \} \}$ and $\cup_{\Sc \in \Dr(A)} \Rr(A, \Sc, X_1,X_2)$ is the set of rate pairs satisfying
\[
R_1 < I(X_1;Y|X_2),
\]
\item $A = \left[\begin{array}{cc}
1 & 0 \\
0 & 0
\end{array} \right]$: $\Dr(A) = \{ \{ 2 \} \}$ and $\cup_{\Sc \in \Dr(A)} \Rr(A, \Sc, X_1,X_2)$ is the set of rate pairs satisfying
\[
R_2 < I(X_2;Y|X_1),
\]
\item $A = \left[\begin{array}{cc}
1 & a \\
0 & 0
\end{array} \right]$ for some nonzero $a \in \Fq$: $\Dr(A) = \{ \{ 1 \}, \{2\} \}$ and
$\cup_{\Sc \in \Dr(A)} \Rr(A, \Sc, X_1,X_2)$ is the set of rate pairs satisfying
\[
R_1 < I(X_1;Y, W_a),
\]
or 
\[
R_2 < I(X_2;Y, W_a),
\]
where $W_a = X_1 \oplus a X_2$.
\end{itemize}
The achievable rate region in Proposition~\ref{prop:shaping_region_k} is then equivalent to $\Rrmac(X_1,X_2) \cap \tRrl(X_1,X_2)$ where $\tRrl(X_1,X_2)$ is the set of rate pairs $(R_1,R_2)$ such that for every nonzero $a$ over $\Fq$
\begin{equation}
R_1 < I(X_1; Y, X_1 \oplus a X_2)
\label{eq:rtl1}
\end{equation}
or
\begin{equation}
R_2 < I(X_2; Y, X_1 \oplus a X_2).
\label{eq:rtl2}
\end{equation}
One may notice that for every nonzero $a$ over $\Fq$
\[
H(X_1 | Y, X_1 \oplus a X_2) = H(X_2 | Y, X_1 \oplus a X_2) \le \min \{ H(X_1 |Y), H(X_2 |Y)  \},
\]
which implies that $\tRrl(X_1,X_2)$ is in general larger than $\Rrl(X_1,X_2)$ defined in Proposition~\ref{prop:shaping_region} in Section~\ref{sec:shaping}. Indeed, the error analysis
in the proof of Proposition~\ref{prop:shaping_region} can be modified to account for the larger
$\tRrl(X_1,X_2)$ region.
\end{remark}

\begin{remark}
The achievable rate region in Proposition~\ref{prop:shaping_region_k} is the largest region thus far established in the literature. As a matter of fact, there is some indication that this region is optimal in the sense that it cannot be improved by using maximum likelihood decoding \cite{Sen--Lim--Kim2018c,Sen--Lim--Kim2018s}. Still, it is in general strictly smaller than the capacity region of the $k$-sender DM-MAC. In particular, for the channels defined in Examples \ref{example:binary_add_mac}--\ref{example:onoff_erasure_mac}, the achievable rate region in Propositon~\ref{prop:shaping_region_k} reduces to the achievable rate region in Proposition~\ref{prop:shaping_region} described in Section~\ref{sec:shaping}. To see this, fix input pmfs $p(x_1)$ and $p(x_2)$. The set of rate pairs satisfying (\ref{eq:rtl1}) or (\ref{eq:rtl2}) for $a=1$ is equivalent to the rate region $\Rrl(X_1,X_2)$.
\end{remark}

As a corollary of Proposition~\ref{prop:shaping_region_k}, we can come up with a smaller rate region achievable by random homologous codes that is easier to compute. As we will discuss in the next section, however, this smaller achievable rate region combined with channel transformation gives rise to the achievability of the capacity region. Let $\Bc$ denote the set of rank deficient $k \times k$ matrices over $\Fq$ that is not row equivalent to a diagonal matrix. Note that $\Bc \subset \Ac$. Given a matrix $A \in \Bc$, a set $\Sc \in \Dr(A)$, and input pmfs $p(x_1),p(x_2),\ldots,p(x_k)$, we define the rate region $\tRr(A, \Sc, X^k)$ as the set of rate tuples
$(R_1,R_2,\ldots,R_k)$ satisfying
\begin{equation*} 
\sum_{j \in \Sc} R_j <  H(X_{\Sc}) - \min_{\Jc \in \Dr(A)} H(X_{\Jc}|Y).
\end{equation*}
Given input pmfs $p(x_1), p(x_2),\ldots,p(x_k)$, we define the rate region 
\begin{equation}
\label{eq:l_shape_k}
\Rrl(X^k) = \bigcap_{A \in \Bc} \; \bigcup_{\Sc \in \Dr(A)} \tRr(A, \Sc, X^k).
\end{equation}

\begin{corollary}
\label{cor:shaping_k}
A rate tuple $(R_1,R_2,\ldots,R_k)$ is achievable for the finite-field input DM-MAC $p(y | x_1,x_2,\ldots,x_k)$ by random homologous codes if
\[
(R_1,R_2,\ldots,R_k) \in \Rrmac(X^k) \cap \Rrl(X^k)
\]
for some input pmfs $p(x_1), p(x_2),\ldots,p(x_k)$.
\end{corollary}

\begin{remark}[Revisit of the $2$-sender DM-MAC with Corollary~\ref{cor:shaping_k}]
For the case $k=2$, the achievable rate region in Corollary~\ref{cor:shaping_k} reduces to the achievable rate region in Proposition~\ref{prop:shaping_region}. To see this, fix the input pmfs $p(x_1)$ and $p(x_2)$. A rank-deficient $2 \times 2$ matrix over $\Fq$ that is not row equivalent to a diagonal matrix must be of the form
\[
\left[ \begin{array}{cc}
a_1 & a_2 \\
0 & 0
\end{array} \right]
\]
for some nonzero $a_1$ and $a_2$ over $\Fq$. Then, for every such matrix $A$, $\Dr(A) = \{ \{1\}, \{2\} \}$. Therefore, the rate region $\Rrl(X_1,X_2)$ defined in \eqref{eq:l_shape_k} is the set of rate pairs $(R_1,R_2)$ such that
\[
R_1 < H(X_1) - \min\{H(X_1|Y), H(X_2|Y)\}
\]
or
\[
R_2 < H(X_2) - \min\{H(X_1|Y), H(X_2|Y)\},
\]
which is equivalent to the rate region $\Rrl(X_1,X_2)$ defined in Section~\ref{sec:shaping}.
\end{remark}

\begin{IEEEproof}[Proof of Corollary~\ref{cor:shaping_k}]
We will show that given input pmfs $p(x_1), p(x_2),\ldots,p(x_k)$ 
\[
(\Rrmac(X^k) \cap \Rrl(X^k)) \: \sbq \: \bigcap_{A \in \Ac} \; \bigcup_{\Sc \in \Dr(A)} \Rr(A, \Sc,X^k),
\]
by first showing that $\Rrmac(X^k) = \bigcap_{A \in \Ac \setminus \Bc} \; \bigcup_{\Sc \in \Dr(A)} \Rr(A, \Sc,X^k)$, and then showing that $\Rrl(X^k) \sbq \bigcap_{A \in \Bc} \; \bigcup_{\Sc \in \Dr(A)} \Rr(A, \Sc,X^k)$. To prove the first claim, let $A$ be a rank-deficient $k\times k$ matrix that is row equivalent to a diagonal matrix $D$ (i.e., $A \in \Ac \setminus \Bc$), and let $\Sc$ be the set of indices such that $i \in \Sc$ if $D_{ii}=0$. Then, by Lemma \ref{lem:pos_indep} in Appendix F, $\Dr(A)=\Sc$ and $\Rr(A, \Sc,X^k)$ is reduced to the set of rate tuples $(R_1,R_2,\ldots,R_k)$ satisfying
\[
\sum_{i \in \Sc} R_{\Sc} < I(X_{\Sc};Y, X_{\Sc^c}).
\]
Taking the intersection over all $A \in \Ac \setminus \Bc$ proves the first claim. For the second claim, it suffices to show that given a matrix $A \in \Bc$ and a set $\Sc \in \Dr(A)$
\[
\tRr(A, \Sc, X^k) \sbq \Rr(A, \Sc, X^k).
\]
Now, a rate tuple $(R_1,R_2,\ldots,R_k) \in \tRr(A, \Sc, X^k)$ satisfies
\begin{align*}
\sum_{j \in \Sc} R_j &<  H(X_{\Sc}) - \min_{\Jc \in \Dr(A)} H(X_{\Jc}|Y) \\
& \le H(X_{\Sc}) - \min_{\Jc \in \Dr(A)} H(X_{\Jc}|Y, W_A) \\
& \overset{(a)}{=} H(X_{\Sc}) - H(X_{\Sc}|Y, W_A), \\
&= I(X_{\Sc};Y, W_A),
\end{align*}
where $(a)$ follows since $H(X_{\Jc}|Y, W_A) = H(X^k|Y,W_A)$ is constant for any $\Jc \in \Dr(A)$. Then, we have $(R_1,R_2,\ldots,R_k) \in \Rr(A, \Sc, X^k)$, which completes the proof.
\end{IEEEproof}

\subsection{Combination}
We incorporate channel transformation with shaping to prove the achievability of the capacity region of the $k$-sender DM-MAC by random homologous codes. 
Similar to the idea discussed in Section~\ref{sec:channel_trans}, we can simply transform the channel $p(y | x_1,x_2,\ldots,x_k)$ into a \emph{virtual channel} with finite-field inputs
\begin{equation} \label{eq:virtual_channel_k}
p(y | u_1,u_2,\ldots,u_k) = p_{Y | X_1,X_2,\ldots,X_k}(y | \vf_1(u_1),\vf_2(u_2),\ldots,\vf_k(u_k))
\end{equation} 
for some symbol-by-symbol mappings $\vf_j: \Fq \to \Xc_j$ for $j \in [1:k]$. Note that this transformation can be applied to any DM-MAC $p(y|x_1,x_2,\ldots,x_k)$ of arbitrary (not necessarily finite-field) input alphabets.

Now, consider the virtual channel in \eqref{eq:virtual_channel_k}
and random homologous codes for this channel. Then, Corollary~\ref{cor:shaping_k} implies the following.

\begin{proposition}
\label{pro:phy_mac_k}
A rate tuple $(R_1,R_2,\ldots,R_k)$ is achievable for the DM-MAC $p(y | x_1,x_2,\ldots,x_k)$
by random homologous codes in $\Fq$,
if
\[
(R_1,R_2,\ldots,R_k) \in \Rrmac(X^k) \cap \Rrl(U^k)
\]
for some $p(u_1),p(u_2),\ldots,p(u_k)$ on $\Fq$ and some mappings 
$x_1 = \vf_1(u_1),x_2 = \vf_2(u_2),\ldots,x_k=\vf_k(u_k)$,
where $\Rrl(U^k)$ is the set of rate tuples $(R_1,R_2,\ldots,R_k)$ satisfying (\ref{eq:l_shape_k}) for the virtual channel $p(y|u_1,u_2,\ldots,u_k)$.
\end{proposition}

We are now ready to extend Theorem~\ref{thm:shaping_cap} to the $k$-sender case, which follows from
Proposition~\ref{pro:phy_mac_k} by optimizing over all channel transformations.

\begin{theorem}
\label{thm:shaping_cap_k}
A rate tuple $(R_1,R_2,\ldots,R_k)$ is achievable for the DM-MAC $p(y | x_1,x_2,\ldots,x_k)$
by random homologous codes in some finite field, if 
\[
(R_1,R_2,\ldots,R_k) \in \Rrmac(X^k)
\]
for some $p(x_1),p(x_2),\ldots,p(x_k)$.
\end{theorem}

\begin{IEEEproof}
We follow similar arguments to the proof of Theorem \ref{thm:shaping_cap}. It suffices to show that given input pmfs $p(x_1),p(x_2),\ldots,p(x_k)$, there exist a finite field $\Fq$, pmfs $p(u_1),p(u_2),\ldots,p(u_k)$ on $\Fq$, and mappings $x_1 = \vf_1(u_1),x_2 = \vf_2(u_2),\ldots,x_k=\vf_k(u_k)$ such that 
\begin{equation}
\Rrmac(X^k) \sbq \Rrl(U^k).
\label{eq:inclusion}
\end{equation}

First, suppose that $p(x_j)$, $j \in [1:k]$, are of the form
$ i/p^m $ for some $i,m \in \Zz^+$ and prime $p$. We consider random homologous codes over $\Fq$ with $q = p^{k^{k}m}$. Let $q_j = p^{k^{(k-j+1)}m}$ for $j \in [1:k]$ and note that 
\[
\Ffi_{q_k} \subset \Ffi_{q_{k-1}} \subset \cdots \subset \Ffi_{q_1} = \Fq.
\]
Consider $U_{j} \sim \U(\Ffi_{q_j})$, and $\vf_{j}$ such that $\vf_{j}(U_{j}) \overset{d}{=} X_{j}$ for $j \in [1:k]$
(this is possible due to the form of $p(x_j)$ and by the choice of $q_j$). To see (\ref{eq:inclusion}), it suffices to show that for every matrix $A \in \Bc$, $\Rrmac(X^k) \sbq \cup_{\Sc \in \Dr(A)} \tRr(A, \Sc, U^k)$. Consider a rate tuple $(R_1,R_2,\ldots,R_k) \in \Rrmac(X^k)$ and a matrix $A \in \Bc$.
By Lemma~\ref{lem:pos_indep} (see Appendix F) and by the choice of $p(u_j)$, there exist at least two different sets $\Sc_1,\Sc_2 \in \Dr(A)$ such that
\[
H(U_{\Sc_1}) - H(U_{\Sc_2}) \ge k \log p^m \ge H(X^k).
\]
Then, $(R_1,R_2,\ldots,R_k)$ satisfies
\begin{align*} 
\sum_{j \in \Sc_1} R_j &< H(X^k) \\
& \le H(U_{\Sc_1}) - H(U_{\Sc_2}) \\
& \le H(U_{\Sc_1}) - \min_{\Jc \in \Dr(A)} H(U_{\Jc}) \\
& \le H(U_{\Sc_1}) - \min_{\Jc \in \Dr(A)} H(U_{\Jc} | Y),
\end{align*}
which implies that $(R_1,R_2,\ldots,R_k) \in \tRr(A, \Sc_1, U^k)$. The claim follows since $A$ is an arbitrary set in $\Bc$. The restrictions on the input pmfs can be removed again by the denseness argument.
\end{IEEEproof}

\subsection{Multiple-receiver Multiple Access Channels}
\label{sec:compound}
Consider the two-receiver DM-MAC $p(y_1,y_2|x_1,x_2)$, where each sender wishes to convey its own message to both of the receivers. Given input pmfs $p(x_1)$ and $p(x_2)$, define $\Rrmac^{(1)}(X_1,X_2)$ as the set of rate pairs satisfying
\begin{align*}
R_1 &\le I(X_1;Y_1|X_2), \\
R_2 &\le I(X_2;Y_1|X_1), \\
R_1+R_2 &\le I(X_1,X_2;Y_1), 
\end{align*}
and $\Rrmac^{(2)}(X_1,X_2)$ as the set of rate pairs satisfying
\begin{align*}
R_1 &\le I(X_1;Y_2|X_2), \\
R_2 &\le I(X_2;Y_2|X_1), \\
R_1+R_2 &\le I(X_1,X_2;Y_2). 
\end{align*}
The following proposition then characterizes the achievable rate region by random homologous codes.
\begin{proposition}
\label{pro:compound}
A rate pair $(R_1,R_2)$ is achievable for the two-receiver DM-MAC $p(y_1,y_2|x_1,x_2)$ by random homologous codes in some finite field, if 
\[
(R_1 ,R_2) \in \Rrmac^{(1)}(X_1,X_2) \cap \Rrmac^{(2)}(X_1,X_2)
\]
for some pmfs $p(x_1)$ and $p(x_2)$.
\end{proposition}

\begin{IEEEproof}
The achievable rate region depends on the conditional pmf $p(y_1,y_2|x_1,x_2)$ only through the conditional marginal pmfs $p(y_1|x_1,x_2)$ and $p(y_2|x_1,x_1)$. First suppose that $p(x_1)$ and $p(x_2)$ are of the form \eqref{eq:prime_pmf}.
We consider random homologous codes over $\Fq$ with $q = p^{2m}$.
Choose $U_1$ and $\vf_1$ such that $U_1$ and $\vf_1(U_1) \overset{d}{=} X_1$
are one-to-one on the support of $U_1$ (this is always possible since $q \ge p^m$).
Also choose $U_2 \sim \U(\Fq)$ and $\vf_2$ such that $\vf_2(U_2) \overset{d}{=} X_2$
(this is possible due to the form of $p(x_2)$). By Proposition~\ref{pro:phy_mac}, the achievable rate region is
\[
\Rrmac^{(1)}(X_1,X_2) \cap \Rrl^{(1)}(U_1,U_2,X_1,X_2) \cap 
\Rrmac^{(2)}(X_1,X_2) \cap \Rrl^{(2)}(U_1,U_2,X_1,X_2),
\]
where $\Rrl^{(j)}(U_1,U_2,X_1,X_2), j=1,2,$ is the set of rate pairs $(R_1,R_2)$ satisfying (\ref{eq:rl1}) or (\ref{eq:rl2}) for the DM-MAC $p(y_j|x_1,x_2)$. The argument in the proof of Theorem \ref{thm:shaping_cap} can be applied to both of the DM-MACs $p(y_1|x_1,x_2)$ and $p(y_2|x_1,x_2)$. As a result, the rate region $\Rrmac^{(j)}(X_1,X_2) \cap \Rrl^{(j)}(U_1,U_2,X_1,X_2), j=1,2,$ is equivalent to the rate region $\Rrmac^{(j)}(X_1,X_2)$, which implies the claim. The restriction on the input pmfs can be removed by the denseness argument.
\end{IEEEproof}

As shown in the examples of the binary adder MAC, the binary erasure MAC, and the on--off
erasure MAC, the insufficiency of shaping or channel transformation for \emph{single-receiver} MACs
can be overcome by time sharing. Indeed, either shaping or channel transformation can achieve
the corner points of $\Rrmac(X_1, X_2)$ of a general DM-MAC $p(y|x_1,x_2)$.
This is no longer the case for multiple receivers, however. 
As illustrated by the following example, 
a proper combination of shaping and channel transformation, even with time sharing, 
can strictly outperform shaping or channel transformation alone.

\begin{example}[A two-receiver MAC]
\label{ex:two_mac}
Let $Y_1 = X_1 + X_2$ (binary erasure MAC), and $Y_2 = (2X_1-1) + Z (2X_2 -1)$ (on--of erasure MAC), where $\Xc_1 = \Xc_2 = \{0,1\}$ and
$Z \sim \Bern(2/3)$ is independent of $X_1$ and $X_2$. The capacity region of this two-receiver MAC is achieved by random coding with i.i.d.\@ $\Bern(1/2)$ inputs $X_1$ and $X_2$, and is sketched in Fig.~\ref{fig:two_rec_cap}. The achievable rate region via shaping in Proposition~\ref{prop:shaping_region} (and Proposition~\ref{prop:shaping_region_k}) is 
\[
\Rrmac^{(1)}(X_1,X_2) \cap \Rrl^{(1)}(X_1,X_2) \cap 
\Rrmac^{(2)}(X_1,X_2) \cap \Rrl^{(2)}(X_1,X_2),
\]
where $\Rrl^{(j)}(X_1,X_2), j=1,2,$ is the set of rate pairs $(R_1,R_2)$ satisfying (\ref{eq:r1}) or (\ref{eq:r2}) for the DM-MAC $p(y_j|x_1,x_2)$, and is shown in Fig.~\ref{fig:two_rec_shaping}. Even after convexification via time sharing, it is strictly smaller than the capacity region with the largest symmetric rate of $11/18$, whereas the symmetric capacity is $2/3$. In comparison, we can combine shaping with channel transformation to achieve the entire capacity region as follows. Consider random homologous codes over $\GF(4) = \{0,1,\a,\a+1\}$. Let $U_1 \sim \U(\GF(4))$ and  $U_2 \sim \Bern(1/2)$ be independent. For channel transformation, let $x_j = \vf(u_j)$ where $\vf(0) = \vf(\a) = 0$, and $\vf(1)=\vf(\a+1)=1$. By this construction, $X_1$ and $X_2$ are i.i.d.\@ $\Bern(1/2)$. Following similar steps to the proof of Proposition~\ref{pro:compound}, it is easy to see that the achievable rate region under this construction is equivalent to $\Rrmac^{(1)}(X_1,X_2) \cap \Rrmac^{(2)}(X_1,X_2)$, which is the capacity region of this channel since $p(x_1)$ and $p(x_2)$ are chosen as the capacity-achieving distributions. 
Thus, combination of shaping with channel transformation not only achieves higher rates than shaping technique, but also achieves the capacity region \emph{without} the need for time sharing.
\begin{figure}[t]
\begin{subfigure}[t]{0.5\textwidth}
\center
\includegraphics[scale=0.7]{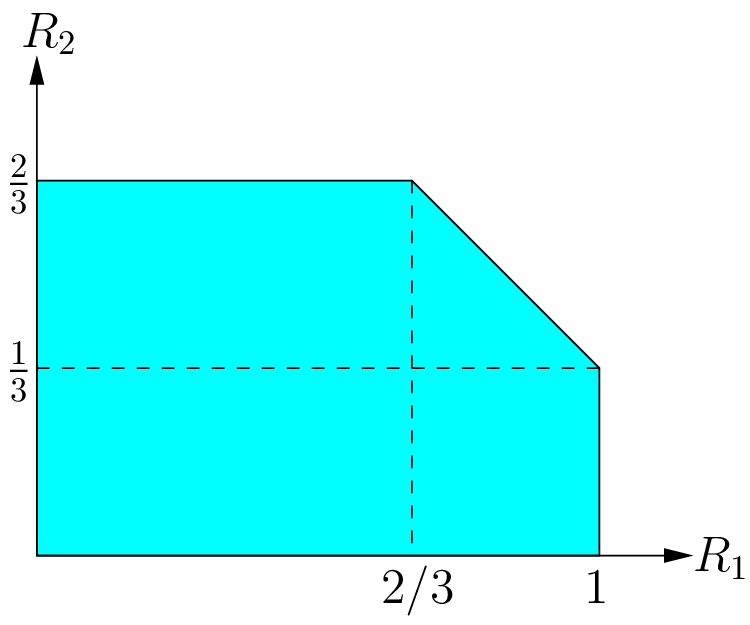}
\caption{}
\label{fig:two_rec_cap}
\end{subfigure}%
\begin{subfigure}[t]{0.5\textwidth}
\center
\includegraphics[scale=0.7]{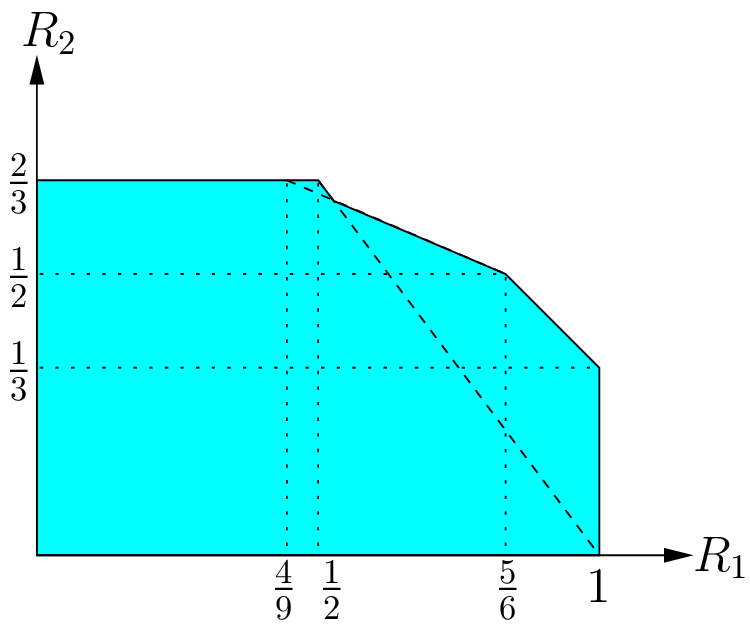}
\caption{}
\label{fig:two_rec_shaping}
\end{subfigure}
\caption{(a) The capacity region and (b) an achievable rate region for the two-receiver MAC.}
\end{figure}
\end{example}

\begin{remark}
Proposition~\ref{pro:compound} can be extended to $k$-sender and $r$-receiver DM-MACs
and compound MACs via the proof of Theorem~\ref{thm:shaping_cap_k}.
\end{remark}

\section{Gaussian Multiple Access Channel}
\label{sec:gauss}
Consider the $2$-sender Gaussian MAC model
\[
Y = g_1 X_{1} + g_2 X_{2} + Z, 
\]
with channel gains $g_1$ and $g_2$, additive noise $Z \sim \N(0,1)$, and average power constraints $\sum_{i=1}^n x_{ji}^2(m_j) \le nP$ for $j=1,2$. Let $S_j = g_j^2 P$, $j=1,2$. The following theorem establishes the achievability of the capacity region of the Gaussian MAC by random homologous codes.

\begin{theorem}
\label{thm:gauss}
A rate pair $(R_1,R_2)$ is achievable for the $2$-sender Gaussian MAC
by random homologous codes, if 
\begin{align*}
R_1 & \le \C(S_1), \\
R_2 & \le \C(S_2), \\
R_1 + R_2 & \le \C(S_1+S_2),
\end{align*}
where $\C(x)= (1/2) \log(1+x), x\ge 0,$ is the Gaussian capacity function.
\end{theorem}

\begin{IEEEproof}
Theorem~\ref{thm:gauss} can be proved using the discretization argument in \cite[Section 3.4.1]{El-Gamal--Kim2011} together with the achievability proof for the $2$-sender DM-MAC by random homologous codes. The proof along this line, however, needs two limit arguments---one for approximating a Gaussian random variable by a discrete random variable, and one for approximating the resulting pmf on a finite alphabet to the desired form in \eqref{eq:prime_pmf}. We instead 
provide a simpler proof via a discretization mapping that results in a pmf of desired form in \eqref{eq:prime_pmf}. 

Let $X_1$ and $X_2$ be i.i.d.\@ $\N(0,P)$. For every $j=1,2,\ldots$, let $[X_1]_j$ be a quantized version of $X_1$ obtained by mapping $X_1$ to the closest point such that $F_{X_1}(X_1) \in [\frac{i}{2^j}, \; \frac{i+1}{2^j}]$ for some positive integer $i$ and $|[X_1]_j| \le |X_1|$, where $F_{X_1}(x)$ denotes the cdf of random variable $X_1$. Clearly, $\E([X_1]_j^2) \le \E(X_1^2) = P$ and the pmf of $[X_1]_j$ is of the form $r/2^j$ for some positive integer $r$. Define $[X_2]_j$ in a similar manner. Let $Y_j = g_1 [X_1]_j + g_2 [X_2]_j + Z$ be the output corresponding to the input pair $[X_1]_j$ and $[X_2]_j$, and let $[Y_j]_k$ be a quantized version of $Y_j$ defined in the same manner. Now, by the achievability proof of Theorem \ref{thm:shaping_cap}, for each $j,k$, random homologous codes over $\Fq$ with $q=2^{2j}$ can achieve the rate pair satisfying
\begin{align*}
R_1 & \le I([X_1]_j;[Y_j]_k|[X_2]_j), \\
R_2 & \le I([X_2]_j;[Y_j]_k|[X_1]_j), \\
R_1 + R_2 & \le I([X_1]_j,[X_2]_j;[Y_j]_k).
\end{align*}
By this type of discretization, weak convergence of $[X_1]_j$ to $X_1$ and $[X_2]_j$ to $X_2$ is preserved, and $([Y_j]_k - Y_j)$ tends to $0$ as $k \to \infty$. Therefore, one can follow the same steps in the proof of \cite[Lemma 3.2]{El-Gamal--Kim2011} to show that 
\begin{align*}
\liminf_{j \to \infty} \lim_{k \to \infty} I([X_1]_j;[Y_j]_k|[X_2]_j) \ge I(X_1;Y|X_2), \\
\liminf_{j \to \infty} \lim_{k \to \infty} I([X_2]_j;[Y_j]_k|[X_1]_j) \ge I(X_2;Y|X_1), \\
\liminf_{j \to \infty} \lim_{k \to \infty} I([X_1]_j,[X_2]_j;[Y_j]_k) \ge I(X_1,X_2;Y),
\end{align*}
which establishes the claim.
\end{IEEEproof}

\begin{remark}
It is straightforward to extend the discretization argument described for the $2$-sender Gaussian MAC to the $k$-sender case. Therefore, random homologous codes can achieve the capacity region of a Gaussian MAC in general. 
\end{remark}

\section{Concluding Remarks}
\label{sec:conc}
In this paper, we examined the possibility of reestablishing the well-known achievable rate regions by random code ensembles for the MACs by using structured, homologous codes. We identified two key techniques to employ nonuniform codewords while preserving a similar structure across the codes of users. The analysis tools developed for these techniques, shaping and channel transformation, imply that their individual performance is insufficient. It is unclear, however, whether there is a fundamental limitation behind each technique. As a constructive alternative to these two techniques and their limits, we showed that an appropriately designed combination of the two can establish the performance of random code ensembles. 
This development and its generalization to multiple senders and receivers motivate
further research into the potential of homologous coding in network information theory.

One immediate question would be whether one can use homologous codes for computation and communication at the same time. To be more specific, consider a multiple-receiver MAC in which some receivers wish to recover a desired linear combination of messages or codewords (computation) while the other receivers wish to recover the message tuples themselves (communication). The definition of computation problem varies in the literature. Some studies, including lattice codes \cite{Nazer--Gastpar2011}, nested coset codes~\cite{Padakandla--Pradhan2013c, Zhu--Gastpar2017}, focus on recovering a desired linear combination of \emph{physical} codewords $X_j^n$, $j = [1:k]$. When the encoding mapping from message to codeword is linear, these two definitions can be used interchangeably. Other studies, including the compute--forward framework recently studied in~\cite{Lim--Gastpar2016}, focus on recovering a desired linear combination of 
\emph{auxiliary} codewords $U_j^n$, $j = [1:k]$, or equivalently, a linear combination of
$[M_1 \; L_1]$ and $[M_2 \; L_2]$ in our notation of homologous codes. This latter setting
simplifies as the computation of a desired linear combination of \emph{messages}, when the rates are symmetric.

The following example demonstrates how random homologous codes discussed thus far
can be adapted for both communication and computation of the messages, highlighting the potential of homologous codes for a broader class of applications beyond multiple access communication. In particular,
random homologous codes, combined with carefully chosen channel transformation,
can achieve rates higher than conventional random codes and homologous codes without channel transformation.

\begin{example}[Simultaneous computation and communication over a multiple-receiver MAC]
\label{ex:compute_communicate}
Consider the $2$-sender and $3$-receiver DM-MAC $p(y_1,y_2,y_3|x_1,x_2)$ where $\Xc_1 = \Xc_2 = \{0,1\}$, and
\begin{align*}
Y_1 & = X_1 \oplus X_2,  && \text{(binary adder MAC)}\\
Y_2 & = X_1 + X_2, && \text{(binary erasure MAC)} \\
Y_3 & = (2X_1-1) + Z (2X_2-1), && \text{(on--of erasure MAC)}
\end{align*}
where $Z\sim \Bern(2/3)$ is independent of $X_1$ and $X_2$. Receiver 1 wishes to recover $M_1 \oplus_q M_2$ over some finite field $\Fq$, whereas both receivers $2$ and $3$ wish to recover the message pair $(M_1,M_2)$. 

First, the optimal achievable rate region by random i.i.d.\@ coding is the intersection of the capacity regions of the DM-MACs $p(y_1|x_1,x_2),p(y_2|x_1,x_2)$, and $p(y_3|x_1,x_2)$, each of which is achieved by i.i.d.\@ $\Bern(1/2)$ inputs $X_1$ and $X_2$, and so is the intersection. Fig.~\ref{fig:compute_comm_iid} sketches the rate region. The optimal symmetric rate for random i.i.d.\@ coding is $1/2$. 

Next, consider binary random homologous codes. By \cite{Lim--Gastpar2016}, given input pmfs $p(x_1)$ and $p(x_2)$, a rate pair $(R_1,R_2)$ is achievable for computing $X_1^n \oplus X_2^n$ (or equivalently, $[M_1 \; L_1] \oplus [M_2 \; L_2]$) at receiver 1 if
\begin{align}
\label{eq:compute1}
R_1 < H(X_1) - H(X_1 \oplus X_2|Y_1) = H(X_1), \\
R_2 < H(X_2) - H(X_1 \oplus X_2|Y_1) = H(X_2).
\label{eq:compute2}
\end{align}
For receivers 2 and 3, a rate pair $(R_1,R_2)$ is achievable (Propositions \ref{prop:shaping_region} and \ref{prop:shaping_region_k}) if  
\begin{equation}
\label{eq:comm1}
(R_1,R_2) \in \Rrmac^{(2)}(X_1,X_2) \cap \Rrl^{(2)}(X_1,X_2) \cap 
\Rrmac^{(3)}(X_1,X_2) \cap \Rrl^{(3)}(X_1,X_2),
\end{equation}
where $\Rrl^{(j)}(X_1,X_2), j=2,3,$ is the set of rate pairs $(R_1,R_2)$ satisfying (\ref{eq:r1}) or (\ref{eq:r2}) for the DM-MAC $p(y_j|x_1,x_2)$. Since the rate constraints in (\ref{eq:compute1}) and (\ref{eq:compute2}) for receiver 1 are looser than those in (\ref{eq:comm1}) for receivers 2 and 3, the resulting achievable rate region is equal to (\ref{eq:comm1}), sketched earlier in Fig.~\ref{fig:two_rec_shaping} for the two-receiver DM-MAC $p(y_2,y_3|x_1,x_2)$. The largest symmetric rate in this region (after convexification via time sharing) is $11/18$. 

Now, we consider random homologous codes over larger finite fields. We need to be more careful for the choice of channel transformation, because we have an additional receiver for the sum of messages rather than the messages themselves. It is easy to see that the construction proposed for Example \ref{ex:two_mac} results in the same rate region as random i.i.d.\@ codes. Therefore, we introduce a better construction here. Let $U_1 \sim \U(\GF(4))$ and 
\[
U_2 = \left \lbrace \begin{array}{ccc}
0 & \textrm{ with probability } \frac{1-\g}{2} \\
1 & \textrm{ with probability } \frac{1-\g}{2} \\
\a & \textrm{ with probability } \frac{\g}{2} \\
\a+1 & \textrm{ with probability } \frac{\g}{2} \\
\end{array} \right.,
\]
be independent, where $\g = H^{-1}(2/3)$. Let $x_j = \vf(u_j)$ where $\vf(0) = \vf(\a) = 0$, and $\vf(1)=\vf(\a+1)=1$. By this construction, $X_1$ and $X_2$ are i.i.d.\@ $\Bern(1/2)$. On the one hand, by \cite{Lim--Gastpar2016} or by (\ref{eq:compute1}) and (\ref{eq:compute2}) with $(U_1,U_2)$ in place of $(X_1,X_2)$, a rate pair $(R_1,R_2)$ is achievable for computing $U_1^n \oplus U_2^n$ (or equivalently, $[M_1 \; L_1] \oplus [M_2 \; L_2]$) at receiver 1 if
\begin{align}
\label{eq:comp3}
R_1 < H(U_1) - H(U_1 \oplus U_2|Y_1) 
= H(U_1) - H(U_1 \oplus U_2) + H(Y) = 1, \\
R_2 < H(U_2) - H(U_1 \oplus U_2|Y_1)
= H(U_2) - H(U_1 \oplus U_2) + H(Y) = 2/3.
\label{eq:comp4}
\end{align}
On the other hand, Proposition~\ref{pro:phy_mac} implies that a rate pair $(R_1,R_2)$ is achievable for communicating messages to receivers 2 and 3, if
\begin{equation}
\label{eq:comm2}
(R_1,R_2) \in \Rrmac^{(2)}(X_1,X_2) \cap \Rrl^{(2)}(U_1,U_2,X_1,X_2) \cap 
\Rrmac^{(3)}(X_1,X_2) \cap \Rrl^{(3)}(U_1,U_2,X_1,X_2),
\end{equation}
where $\Rrl^{(j)}(U_1,U_2,X_1,X_2), j=2,3,$ is the set of rate pairs $(R_1,R_2)$ satisfying (\ref{eq:rl1}) or (\ref{eq:rl2}) for the DM-MAC $p(y_j|x_1,x_2)$ with specified $p(u_1,x_1)$ and $p(u_2,x_2)$. The achievable rate region consisting of all rate pairs satisfying (\ref{eq:comp3})-(\ref{eq:comm2}) after convexification via time sharing is sketched in Fig.~\ref{fig:compute_comm_cap}. The largest symmetric rate is $2/3$, which can be shown to be the symmetric capacity for this example. Therefore, with the help of channel transformation, the symmetric capacity is achieved by homologous codes.

\begin{figure}[t]
\begin{subfigure}[t]{0.5\textwidth}
\center
\includegraphics[scale=0.7]{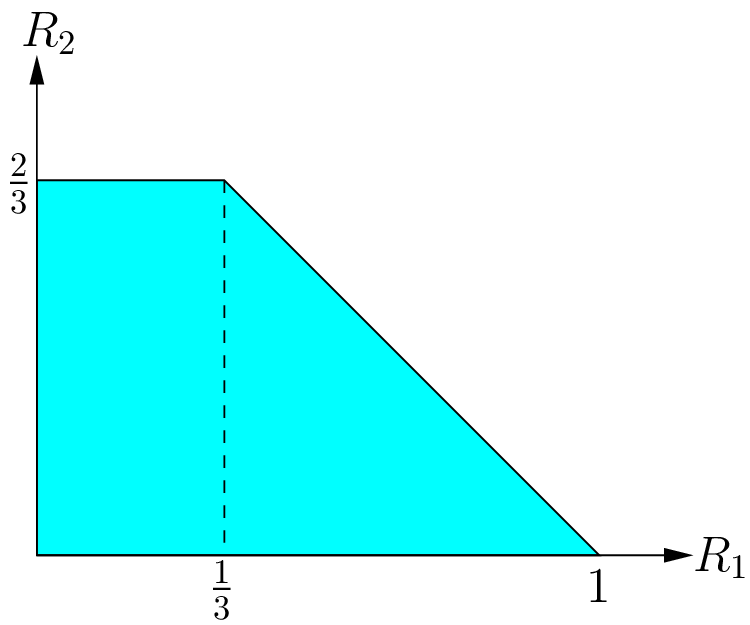}
\caption{}
\label{fig:compute_comm_iid}
\end{subfigure}%
\begin{subfigure}[t]{0.5\textwidth}
\center
\includegraphics[scale=0.7]{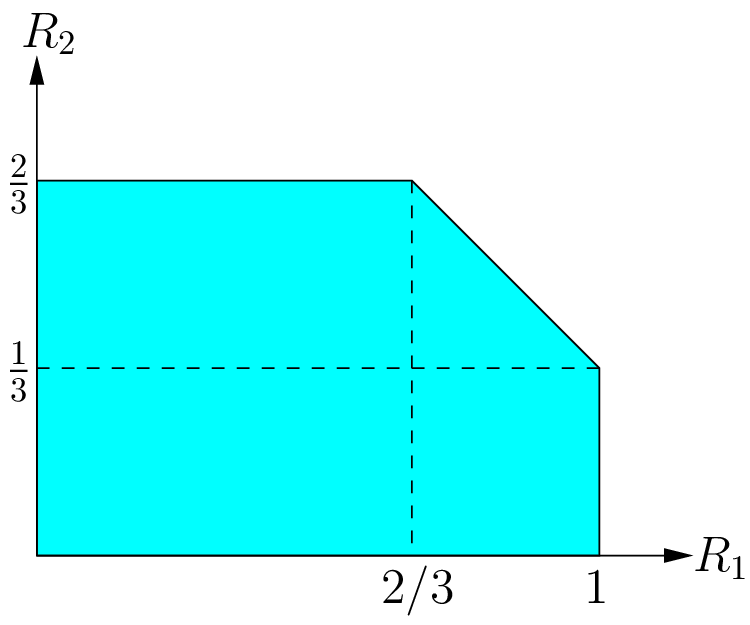}
\caption{}
\label{fig:compute_comm_cap}
\end{subfigure}
\caption{(a) The optimal achievable rate region by random i.i.d.\@ codes and (b) an achievable rate region of the compute-communicate MAC.}
\end{figure}

\end{example}

\section*{Acknowledgments}
This work was supported in part by the Electronics and Telecommunications Research Institute
through Grant 17ZF1100 from the Korean Ministry of Science, ICT, and Future Planning. We would like to thank Dr. Keunyoung Kim for his invaluable comments and questions that prodded us to gain important insights about the problem. 

\section*{Appendices}

\section*{Appendix A}
\begin{claim}
For the binary erasure MAC, no pair of binary coset codes with the same generator matrix can achieve the rate pair $(1/2 + \e, 1/2 + \e)$ for $\e > 0$. 
\end{claim}

\begin{IEEEproof}
Let $\e>0$ and $R_1 = R_2 = R = 1/2+\e$. Suppose without loss of generality that $nR \in \Zz^+$, and that the generator matrix $G$ is a full rank $nR \times n$ matrix and does not have an all zero column. The messages $M_1$ and $M_2$ are assumed to be i.i.d.\@ $\U(\GF(2)^{nR})$. The received sequence is then written as
\[
Y^n = (M_1 G \oplus d_1^n) + (M_2 G \oplus d_2^n).
\]
Define $\Yt_i = (Y_i)\mod 2$ for $i=1,2,\ldots,n$, which implies
\[ \Yt^n = (M_1 \oplus M_2) G  \oplus (d_1^n \oplus d_2^n).\]
Define random set 
$\Sc(\Yt^n) = \{ i: \Yt_i = 0 \}$,
and let random variable $N_0 = | \Sc(\Yt^n) |$ denote the number of positions where sequence $\Yt^n$ has $0$. 
We construct a new (random) matrix $G_{\Sc}$ of size $nR \times N_0$ by including the columns $g_i$ of $G$ for $i \in \Sc$. Then, the decoder makes an error if the following event occurs
\[\Ec = \{ N_0 < nR \}.\]
This observation follows from the fact that on $\Ec$, dimension of null space of $G_{\Sc}$ is strictly larger than $0$, so  $\exists \: (m_1,m_2) \neq (M_1,M_2)$ such that $(m_1 \oplus M_1)G_{\Sc} = \Ob$ and ${m}_1 \oplus {m}_2 = {M}_1 \oplus {M}_2$, which leads to the same received sequence $Y^n$.


By the union of events bound, we have 
$P_e^{(n)} \ge \P(\Ec) = 1 - \P(\Ec^c)$. To bound the probability $\P(\Ec^c)$, we define the coset code $\Cc = \{ x^n \in \GF(2)^{n}: x^n = m G \oplus d_1^n \oplus d_2^n, \; m \in \GF(2)^{nR} \}$. Then, $\Yt^n$ is uniformly distributed among $\Cc$, and we have 
\begin{align*}
P(\Ec^c)  & \overset{(a)}{\leq} \frac{\E [N_0] }{nR} \\
&= \frac{\sum\limits_{x^n \in \Cc} \P(\Yt^n = x^n) wt((x^n)^c)}{nR} \\
&= \frac{\sum\limits_{x^n \in \Cc} 2^{-nR} wt((x^n)^c)}{nR}, \\
& \overset{(b)}{=} \frac{2^{-nR} (n 2^{nR-1})}{nR},\\
& = \frac{1}{1+2 \e},
\end{align*}
where function $wt: \GF(2)^n \rightarrow \Zz^+$ returns the Hamming weight of the input, $(a)$ follows from the Markov inequality and $(b)$ follows from the fact that for a binary coset code $\mathcal{C}$, at a given index, exactly half of the codewords have $0$ and exactly half of the codewords have $1$ (remember that its generator matrix $G$ has no all-zero column). It follows that $ P_e^{(n)} \geq \frac{2\e}{1+2\e}$, which proves the claim.
\end{IEEEproof}

\section*{Appendix B}
\begin{claim}
\label{claim:reg}
Given input pmfs $p(x_1)$ and $p(x_2)$, let $\Rr(X_1,X_2)$ be the rate region that consists of the set of rate pairs $(R_1,R_2)$ such that
\[
\min \{R_1 + H(X_2), R_2 + H(X_1) \} < H(X_1) + H(X_2) - \min \{H(X_1|Y), H(X_2|Y) \}.
\]
The region $\Rr(X_1,X_2)$ is equivalent to the rate region $\Rrl(X_1,X_2)$ described in (\ref{eq:r1}) and (\ref{eq:r2}).
\end{claim}
\begin{IEEEproof}
It is easy to see that $\Rr(X_1,X_2) \sbq \Rrl(X_1,X_2)$. For the other direction, let rate pair $(R_1,R_2) \in \Rrl(X_1,X_2)$ such that $R_1 + H(X_2) \le R_2 + H(X_1)$. By   the definition of the rate region $\Rrl(X_1,X_2)$, we have
\[
R_1 + H(X_2) \le \max \{ H(X_2) + I(X_1;Y), H(X_1) + I(X_2;Y) \},
\]
which implies $(R_1,R_2) \in \Rr(X_1,X_2)$. Similarly, rate pair $(R_1,R_2) \in \Rrl(X_1,X_2)$ such that $R_2 + H(X_1) \le R_1 + H(X_2)$ is in $\Rr(X_1,X_2)$. Therefore, $\Rrl(X_1,X_2) \sbq \Rr(X_1,X_2)$, from which the claim follows.
\end{IEEEproof}

\section*{Appendix C \\ Proposition~\ref{prop:shaping_region} for binary adder MAC}

When specialized to the binary adder MAC, the achievable rate region in Proposition~\ref{prop:shaping_region} is reduced to the rate pairs
$(R_1,R_2)$ such that
\begin{align*} 
R_1 &< I(X_1;Y),\\
R_2 &< I(X_2;Y|X_1) = H(X_2),
\end{align*}
or 
\begin{align*} 
R_1 &< I(X_1;Y|X_2) = H(X_1),\\
R_2 &< I(X_2;Y),
\end{align*}
for some input pmfs $p(x_1)$ and $p(x_2)$, which is equivalent to the capacity region depicted in Fig.~\ref{fig:binary_add_mac_cap}. To see this, let $\a \in [0,1/2]$, and consider $X_1 \sim \Bern (\a)$ and $X_2 \sim \Bern \left(\frac{1}{2} \right)$. Then, the rate pairs $(R_1,R_2)$ which satisfies 
\begin{align*}
R_1 &< H(\a), \\
R_2 &< 1-H(\a)
\end{align*}
is achievable. Since $H(\a)$ is continuous on $\a$, taking the union over $\a \in [0,1/2]$ implies that every point within the capacity region is achievable by shaping technique. It follows from the converse proof for the capacity region of binary adder MAC that the achievable rate region in Proposition~\ref{prop:shaping_region} (over all input pmfs) is indeed equivalent to the capacity region.

\section*{Appendix D \\ binary erasure MAC}
\subsection*{The achievable rate region by shaping}
For the binary erasure MAC, we will evaluate the rate region in Proposition~\ref{prop:shaping_region}. Let $\a,\b \in [0,1/2]$, and consider $X_1 \sim \Bern (\a)$ and $X_2 \sim \Bern (\b)$. By Proposition~\ref{prop:shaping_region}, the set of rate pairs $(R_1,R_2)$ such that 
\begin{align*}
R_1 &< I(X_1;Y) = f(\a,\b), \\
R_2 &< I(X_2;Y|X_1) = H(\b), 
\end{align*}
or 
\begin{align*}
R_1 &< I(X_1;Y|X_2) = H(\a), \\
R_2 &< I(X_2;Y) = f(\b,\a), 
\end{align*}
is achievable, where function $f: [0,1/2] \times [0,1/2] \rightarrow \Real$ is defined as
\begin{align}
\label{eqn:function_f_example}
f(x,y) = H(x)- y (1-x) \log \left(1+\frac{x}{1-x}\,\frac{1-y}{y} \right)
-x (1-y) \log \left(1+\frac{1-x}{x}\,\frac{y}{1-y} \right).
\end{align}
Since $f(x,y)$ is increasing on $x$ for any $y \in [0,1/2]$, the union of such regions over $\a, \b \in[0,1/2]$ is the set of rate pairs $(R_1,R_2)$ satisfying
\begin{align*}
R_1 &< 1 - \frac{H(\a)}{2}, \\
R_2 &< H(\a), 
\end{align*}
or 
\begin{align*}
R_1 &< H(\a), \\
R_2 &< 1 - \frac{H(\a)}{2}, 
\end{align*}
for some $\a \in [0 \; 1/2]$. By the fact that $H(\a) \in [0\; 1]$ is continuous on $\a$, this union is equivalent to the union of two trapezoids defined by
\begin{align*}
R_2 & < 1, \\
2R_1 + R_2 & < 2,
\end{align*}
and
\begin{align*}
R_1 & < 1, \\
R_1 + 2R_2 & < 2,
\end{align*}
which proves the claim.

\subsection*{The achievable rate region by channel transformation}
We will evaluate the achievable rate region by the channel transformation technique for binary erasure MAC. Let $\a,\b \in [0 \; 1/2]$, and consider $X_1 \sim \Bern (\a)$ and $X_2 \sim \Bern (\b)$. By Corollary~\ref{cor:phy_mac_coset}, the set of rate pairs $(R_1,R_2)$ such that
\begin{align} \notag
R_1 {}& < \min\{I(X_1;Y|X_2), \max[I(X_1;Y), I(X_2;Y)] \} = \min\{H(\a), \max[f(\a,\b), f(\b,\a)] \}, \\
R_2 {}& < I(X_2;Y|X_1) = H(\b), 
\label{reg:be_mac_ct1}\\ \notag
\begin{split}
R_1 + R_2 {}& < I(X_1,X_2;Y) = H(\a)+ f(\b,\a) = H(\b) + f(\a,\b),
\end{split}
\end{align}
or
\begin{align}\notag
R_1 {}& < I(X_1;Y|X_2) = H(\a), \\
R_2 {}& <  \min\{I(X_2;Y|X_1), \max[I(X_1;Y), I(X_2;Y)] \} = \min\{H(\b), \max[f(\a,\b), f(\b,\a)] \}, 
\label{reg:be_mac_ct2}\\ \notag
\begin{split}
R_1 + R_2 {}& < I(X_1,X_2;Y) = H(\a)+ f(\b,\a) = H(\b) + f(\a,\b),
\end{split}
\end{align}
is achievable, where function $f$ is as defined in (\ref{eqn:function_f_example}). First, consider the union of such regions over $\a,\b \in [0,1/2]$ such that $\a \geq \b$ (or equivalently $f(\a,\b) \geq f(\b,\a)$), which results in the rate region defined by  
\begin{align*}
R_1 &< f(\a,\b), \\
R_2 &< H(\b), 
\end{align*}
or 
\begin{align*}
R_1 &< H(\a), \\
R_2 &< \min \{ H(\b), f(\a,\b)\},\\
R_1 + R_2 &< H(\b) + f(\a,\b),
\end{align*}
for some $\a,\b \in [0 \; 1/2]$ such that $\a \geq \b$. Since $f(x,y)$ is increasing over $x$ for any $y \in [0 \; 1/2]$, the resulting region consists of the rate pairs $(R_1,R_2)$ satisfying 
\begin{align} 
\label{reg:be_mac_1}
R_1 &< f(1/2,\b) = 1-\frac{H(\b)}{2}, \\ \notag
R_2 &< H(\b), 
\end{align}
or 
\begin{align}\notag
\label{reg:be_mac_2}
R_1 &< 1, \\
R_2 &< \min \{ H(\b), 1-\frac{H(\b)}{2} \},\\ \notag
R_1 + R_2 &< 1+\frac{H(\b)}{2},
\end{align}
for some $\b \in [0,1/2]$. The union of the rate region defined in (\ref{reg:be_mac_1}) over $\b \in [0 \; 1/2]$ is equivalent to the trapezoid defined by $R_2 < 1$, and $2R_1 + R_2 < 2$. The union of the rate region defined in (\ref{reg:be_mac_2}) over $\b \in [0 \; 1/2]$ is clearly included in the trapezoid defined by $R_1 < 1$, $R_1 + 2R_2 < 2$.

By similar arguments, the union of the rate region defined in (\ref{reg:be_mac_ct1}) and (\ref{reg:be_mac_ct2}) over $\a,\b \in [0,1/2]$ such that $\b \geq \a$, is reduced to the rate pairs $(R_1,R_2)$ such that
\begin{align*}
R_1 &< \min \{ H(\a), 1-\frac{H(\a)}{2} \},\\
R_2 &< 1, \\
R_1 + R_2 &< 1+\frac{H(\a)}{2},
\end{align*}
or
\begin{align*}
R_1 &< H(\a),\\ 
R_2 &< 1-\frac{H(\a)}{2}, 
\end{align*}
for some $\a \in [0,1/2]$. By symmetry, the overall achievable rate region in Corollary~\ref{cor:phy_mac_coset} is equivalent to the union of two trapezoids defined by $R_2 < 1$, $2R_1 + R_2 < 2$ and $R_1 < 1$, $R_1 + 2R_2 < 2$.

\section*{Appendix E \\ on--off erasure MAC}
\subsection*{The achievable rate region by shaping}
We will evaluate the achievable rate region described in Proposition~\ref{prop:shaping_region} for on--off erasure MAC. If the channel parameter $p \le 2/3$, it is easy to see that i.i.d.\@ $\Bern(1/2)$ inputs $X_1$ and $X_2$ can achieve the capacity region in Fig.~\ref{fig:onoff_mac_cap}. Suppose that $p > 2/3$. Let $\a,\b \in [0 \; 1/2]$, and consider $X_1 \sim \Bern (\a)$ and $X_2 \sim \Bern (\b)$. Then, by Proposition~\ref{prop:shaping_region}, the set of rate pairs $(R_1,R_2)$  such that
\begin{align}
\label{eq:oo1}
R_1 &< I(X_1;Y) = (1-p)H(\a) + p f(\a,\b), \\
R_2 &< I(X_2;Y|X_1) = pH(\b),
\label{eq:oo2}
\end{align}
or
\begin{align}
\label{eq:oo3}
R_1 &< I(X_1;Y|X_2) = H(\a), \\ \notag
R_2 &< \min\{I(X_2;Y|X_1), H(X_2)-H(X_1)+I(X_1;Y)\}
= \min \{pH(\b), (1-p)H(\b)+pf(\b,\a)\}, \\
R_1 + R_2 &< H(\a) + p f(\b,\a),
\label{eq:oo4}
\end{align}
is achievable, where function $f$ is as defined in (\ref{eqn:function_f_example}). First, consider the union of the rate region defined in (\ref{eq:oo1}) and (\ref{eq:oo2}) over $\a, \b \in [0 \; 1/2]$. Since $f(x,y)$ is increasing on $x$ for every $y \in [0 \; 1/2]$, the union is equivalent to the set of rate pairs $(R_1,R_2)$ satisfying
\begin{align*}
R_1 &< 1 - p + p\left(1 - \frac{H(\b)}{2} \right) = 1- \frac{p H(\b)}{2}, \\
R_2 &< pH(\b), 
\end{align*}
for some $b \in [0 \; 1/2]$, that reduces to the trapezoid defined by $R_2 < p$ and $2R_1+R_2<2$. 

Second, we consider the union of the rate region defined in (\ref{eq:oo3})-(\ref{eq:oo4}) over $\a, \b \in [0 \; 1/2]$. By similar arguments, the union is equivalent to the set of rate pairs $(R_1,R_2)$ such that
\begin{align*}
R_1 &< H(\a), \\ 
R_2 &< \min \{p, 1 - \frac{p H(\a)}{2}\}, \\
R_1 + R_2 &< p + H(\a)\left( 1 - \frac{p}{2} \right),
\end{align*}
for some $\a \in [0 \; 1/2]$, that is equivalent to the hexagon defined by $R_1 < 1$, $R_2 < p$, $R_1 + R_2 < 1+p/2$, and $(p/2)R_1 + R_2 < 1-(p/2) + (p^2)/2$.

\subsection*{The achievable rate region by channel transformation}
We will evaluate the achievable rate region described in Corollary~\ref{cor:phy_mac_coset} for on--off erasure MAC. Again, if the channel parameter $p \le 2/3$, it is easy to see that i.i.d.\@ $\Bern(1/2)$ inputs $X_1$ and $X_2$ can achieve the capacity region in Fig.~\ref{fig:onoff_mac_cap2}. Suppose that $p > 2/3$. Let $\a,\b \in [0 \; 1/2]$, and consider $X_1 \sim \Bern (\a)$ and $X_2 \sim \Bern (\b)$. Then, by Corollary~\ref{cor:phy_mac_coset}, the set of rate pairs $(R_1,R_2)$  such that
\begin{align}
\label{eq:oo_ct1}
R_1 &< I(X_1;Y|X_2) = H(\a), \\ \notag
R_1 &< \max\{I(X_1;Y), I(X_2;Y)\}
= \max \{ p f(\a,\b)+(1-p)H(\a), p f(\b,\a)\}, \\ \notag
R_2 &< I(X_2;Y|X_1) = p H(\b), \\ 
R_1 + R_2 &< I(X_1,X_2;Y) = H(\a) + p f(\b,\a),
\label{eq:oo_ct2}
\end{align}
or
\begin{align}
\label{eq:oo_ct3}
R_1 &< I(X_1;Y|X_2) = H(\a), \\ \notag
R_2 &< I(X_2;Y|X_1) = p H(\b), \\ \notag
R_2 &< \max\{I(X_1;Y), I(X_2;Y)\}
= \max \{ p f(\a,\b)+(1-p)H(\a), p f(\b,\a)\}, \\
R_1 + R_2 &< I(X_1,X_2;Y) = H(\a) + p f(\b,\a),
\label{eq:oo_ct4}
\end{align}
is achievable, where function $f$ is as defined in (\ref{eqn:function_f_example}). First, consider the union of the rate region defined in (\ref{eq:oo_ct1})-(\ref{eq:oo_ct2}) over $\a, \b \in [0 \; 1/2]$ such that $H(\a) > pH(\b)$ (or equivalently $p f(\a,\b)+(1-p)H(\a) > p f(\b,\a)$). Then, the inequalities in (\ref{eq:oo_ct1}) and (\ref{eq:oo_ct2}) are inactive. Since $f(x,y)$ is increasing on $x$ for every $y \in [0 \; 1/2]$, the union is equivalent to the set of rate pairs $(R_1,R_2)$ satisfying
\begin{align*}
R_1 &< p\left(1 - \frac{H(\b)}{2} \right) +(1-p)
= 1- \frac{p H(\b)}{2}, \\
R_2 &< pH(\b), 
\end{align*}
for some $\b \in [0 \; 1/2]$, that reduces to the trapezoid defined by $R_2 < p$ and $2R_1+R_2<2$. It is easy to see that the union of the rate region defined in (\ref{eq:oo_ct1})-(\ref{eq:oo_ct2}) over $\a, \b \in [0 \; 1/2]$ such that $H(\a) \le pH(\b)$ is included in this trapezoid.

Second, we consider the union of the rate region defined in (\ref{eq:oo_ct3})-(\ref{eq:oo_ct4}) over $\a, \b \in [0 \; 1/2]$ such that $H(\a) > pH(\b)$. By similar arguments, the union is equivalent to the set of rate pairs $(R_1,R_2)$ such that
\begin{align*}
R_1 &< 1, \\ 
R_2 &< \min \{pH(\b), 1 - \frac{p H(\b)}{2}\}, \\
R_1 + R_2 &< 1 + \frac{p}{2} H(\b),
\end{align*}
for some $\b \in [0 \; 1/2]$, that is equivalent to the hexagon defined by $R_1 < 1$, $R_2 < 2/3$, $R_1 + R_2 < 1+p/2$, and $R_1 + 2 R_2 < 2$. Finally, it is easy to see that the union of the rate region defined in (\ref{eq:oo_ct3})-(\ref{eq:oo_ct4}) over $\a, \b \in [0 \; 1/2]$ such that $H(\a) \le pH(\b)$ is equivalent to the trapezoid defined by $R_1 < p$ and $(p/2)R_1 + R_2 < p$.

\section*{Appendix F}

\begin{lemma}
\label{lem:pos_indep}
Suppose that $Z = \{z_1,z_2,\ldots,z_r\}$ is a set of linearly independent vectors in a vector space $V$ of dimension $k>r$, and $W = \{w_1,w_2,\ldots,w_k\}$ span $V$. Let $T \sbq W$ be a set such that

i) $|T|=k-r$, and

ii) $Z \cup T$ span $V$. 

\noindent (The existence of such $T$ is guaranteed by the Steinitz Lemma in~\cite{Katznel--Katznel2008}). Then, $T = W \setminus J$ is the unique subset of $W$ satisfying i) and ii) if and only if $span(Z) = span(J)$, where $J \sbq W$ with $|J|=r$. 
\end{lemma}

\begin{IEEEproof}
Let $J \sbq W$ with $|J|=r$. First suppose that $span(Z) = span(J)$. Then, it is easy to see that $T = W \setminus J$ is the only subset of $W$ that satisfies i) and ii). Now, suppose that
$T=W \setminus J$ is the unique subset of $W$ that satisfies i) and ii). We will show that 
\[
span(Z) = span(J).
\] 
Both $Z$ and $J$ consist of $r$ linearly independent vectors, so it suffices to show that for every $w \in J$, $w \in span(Z)$. Let $w \in J$. Since $Z \cup T$ span $V$, we have
\begin{equation}
w = \sum_{l=1}^r a_l z_l + \sum_{w_i \in T} b_i w_i.
\label{eq:unique}
\end{equation}
We want to show that $b_i =0$ for all $w_i \in T$ in (\ref{eq:unique}). Assume to the contrary that $b_m \neq 0$ for some $w_m \in T$. Then we can write $w_m$ as a linear combination of the vectors in $Z \cup T \setminus \{w_m\} \cup \{w \}$. Note that $w \neq w_m$ since $J$ and $T$ are disjoint. Thus, $T' \triangleq  T \setminus \{w_m\} \cup \{w \}$ also satisfies i) and ii), which contradicts with the uniqueness of $T$. The claim follows since $w \in J$ is arbitrary.
\end{IEEEproof}

\bibliography{nit}

\end{document}

%% file: defns.tex
\usepackage{xspace}
\usepackage{bbm}
\input{mathlig}

\newcommand{\muspace}{\mspace{1mu}}

\DeclareRobustCommand{\scond}{\mathchoice{\muspace\vert\muspace}{\vert}{\vert}{\vert}}
\mathlig{|}{\scond}

\DeclareRobustCommand{\discint}{\mathchoice{\mspace{-1.5mu}:\mspace{-1.5mu}}{\mspace{-1.5mu}:\mspace{-1.5mu}}{:}{:}}
\mathlig{::}{\discint}
\newcommand{\suchthat}{\colon}

\def\rank{\mathop{\rm rank}\nolimits}%
\newcommand{\Ac}{\mathcal{A}}
\newcommand{\Bc}{\mathcal{B}}
\newcommand{\Cc}{\mathcal{C}}

\newcommand{\Dc}{\mathcal{D}}
\newcommand{\Ec}{\mathcal{E}}

\newcommand{\Jc}{\mathcal{J}}

\newcommand{\Lc}{\mathcal{L}}

\newcommand{\Sc}{\mathcal{S}}

\newcommand{\Xc}{\mathcal{X}}
\newcommand{\Yc}{\mathcal{Y}}

\newcommand{\Rr}{\mathscr{R}}

\newcommand{\aep}{{\mathcal{T}_{\epsilon}^{(n)}}}
\newcommand{\aepvar}{{\mathcal{T}_{\epsilon'}^{(n)}}}

\newcommand{\Rh}{{\hat{R}}}

\newcommand{\kh}{{\hat{k}}}

\newcommand{\mh}{{\hat{m}}}

\newcommand{\Yt}{{\tilde{Y}}}

\newcommand{\xt}{{\tilde{x}}}

\def\a{\alpha}
\def\b{\beta}
\def\g{\gamma}

\def\d{\delta}
\def\e{\epsilon}

\DeclareMathOperator\E{\textsf{E}}
\let\P\relax
\DeclareMathOperator\P{\textsf{P}}

\DeclareMathOperator\C{\textsf{C}}

\newcommand{\Bern}{\mathrm{Bern}}

\newcommand{\N}{\mathrm{N}}
\newcommand{\U}{\mathrm{Unif}}

\def\textiid{i.i.d.\@\xspace}
\newcommand\iid{\ifmmode\text{ i.i.d. } \else \textiid \fi}

\newcommand{\Zz}{\mathbb{Z}}
\newcommand{\Real}{\mathbb{R}}

\def\clap#1{\hbox to 0pt{\hss#1\hss}}
\def\mathclap{\mathpalette\mathclapinternal}
\def\mathclapinternal#1#2{%
  \clap{$\mathsurround=0pt#1{#2}$}}

\let\oldstackrel\stackrel
\renewcommand{\stackrel}[2]{\oldstackrel{\mathclap{#1}}{#2}}



%% file: mathlig.tex
\catcode`@=11
\chardef\mathlig@atcode\count255

\def\actively#1#2{\begingroup\uccode`\~=`#2\relax\uppercase{\endgroup#1~}}
\def\mathlig@gobble{\afterassignment\mathlig@next@cmd\let\mathlig@next= }
\def\mathlig@delim{\mathlig@delim}
\def\mathlig@defcs#1{\expandafter\def\csname#1\endcsname}
\def\mathlig@let@cs#1#2{\expandafter\let\expandafter#1\csname#2\endcsname}
\def\mathlig@appendcs#1#2{\expandafter\edef\csname#1\endcsname{\csname#1\endcsname#2}}

\def\mathlig#1#2{\mathlig@checklig#1\mathlig@end\mathlig@defcs{mathlig@back@#1}{#2}\ignorespaces}

\def\mathlig@checklig#1#2\mathlig@end{%
 \expandafter\ifx\csname mathlig@forw@#1\endcsname\relax
 \expandafter\mathchardef\csname mathlig@back@#1\endcsname=\mathcode`#1%
 \mathcode`#1"8000\actively\def#1{\csname mathlig@look@#1\endcsname}%
 \mathlig@dolig#1\mathlig@delim
\fi
\mathlig@checksuffix#1#2\mathlig@end
}

\def\mathlig@checksuffix#1#2\mathlig@end{%
\ifx\mathlig@delim#2\mathlig@delim\relax\else\mathlig@checksuffix@{#1}#2\mathlig@end\fi
}
\def\mathlig@checksuffix@#1#2#3\mathlig@end{%
\expandafter\ifx\csname mathlig@forw@#1#2\endcsname\relax\mathlig@dosuffix{#1}{#2}\fi
\mathlig@checksuffix{#1#2}#3\mathlig@end
}

\def\mathlig@dosuffix#1#2{%
\mathlig@appendcs{mathlig@toks@#1}{#2}%
\mathlig@dolig{#1}{#2}\mathlig@delim
}

\def\mathlig@dolig#1#2\mathlig@delim{%
 \mathlig@defcs{mathlig@look@#1#2}{%
 \mathlig@let@cs\mathlig@next{mathlig@forw@#1#2}\futurelet\mathlig@next@tok\mathlig@next}%
 \mathlig@defcs{mathlig@forw@#1#2}{%
  \mathlig@let@cs\mathlig@next{mathlig@back@#1#2}%
  \mathlig@let@cs\checker{mathlig@chck@#1#2}%
  \mathlig@let@cs\mathligtoks{mathlig@toks@#1#2}%
  \expandafter\ifx\expandafter\mathlig@delim\mathligtoks\mathlig@delim\relax\else
  \expandafter\checker\mathligtoks\mathlig@delim\fi
  \mathlig@next
 }%
 \mathlig@defcs{mathlig@toks@#1#2}{}%
 \mathlig@defcs{mathlig@chck@#1#2}##1##2\mathlig@delim{%
  \ifx\mathlig@next@tok##1%
   \mathlig@let@cs\mathlig@next@cmd{mathlig@look@#1#2##1}\let\mathlig@next\mathlig@gobble
  \fi 
  \ifx\mathlig@delim##2\mathlig@delim\relax\else
   \csname mathlig@chck@#1#2\endcsname##2\mathlig@delim
  \fi
 }%
 \ifx\mathlig@delim#2\mathlig@delim\else
  \mathlig@defcs{mathlig@back@#1#2}{\csname mathlig@back@#1\endcsname #2}%
 \fi
}%

\catcode`@\mathlig@atcode